\pdfoutput=1 
\documentclass[twoside]{IEEEtran}
\usepackage[noadjust]{cite}
\ifCLASSINFOpdf
   \usepackage[pdftex]{graphicx}
\else
\fi
\usepackage{amsmath}
\usepackage{array}
  \usepackage[caption=false,font=footnotesize]{subfig}
\usepackage{url}
\usepackage{multirow}
\usepackage{balance}
\usepackage{epstopdf}
\begin{document}
\title{Speeding up VP9 Intra Encoder with  Hierarchical Deep Learning Based Partition Prediction}
\author{Somdyuti Paul,
        Andrey Norkin,
        and Alan C. Bovik

\thanks{S. Paul and A. C. Bovik are with the Department
of Electrical and Computer Engineering, University of Texas at Austin, Austin,
TX, 78712 USA (email: somdyuti@utexas.edu, bovik@ece.utexas.edu).}% <-this % stops a space
\thanks{A. Norkin is with Netflix Inc. Los Gatos, CA, 95032 USA (email: anorkin@netflix.com).}% <-this % stops a space
\thanks{This work is supported by Netflix Inc.}}

\IEEEpubid{\begin{minipage}{\textwidth}\ \\[12pt] \\
\copyright 2020 IEEE. Personal use of this material is permitted.  Permission from IEEE must be obtained for all other uses, in any current or future media, including reprinting/republishing this material for advertising or promotional purposes, creating new collective works, for resale or redistribution to servers or lists, or reuse of any copyrighted component of this work in other works.”\\ 
\end{minipage}} 


\maketitle

\begin{abstract}
In VP9 video codec, the sizes of blocks are decided during encoding by recursively partitioning 64$\times$64 superblocks using rate-distortion optimization (RDO). This process is computationally intensive because of the  combinatorial search space of possible partitions of a superblock. Here, we propose a deep learning based alternative framework to predict the intra-mode superblock partitions in the form of a four-level partition tree, using a hierarchical fully convolutional network (H-FCN). We created a large database of VP9 superblocks and the corresponding partitions to train an H-FCN model, which was subsequently integrated with the VP9 encoder to reduce the intra-mode encoding time. The experimental results establish that our approach speeds up intra-mode encoding by 69.7\% on average, at the expense of a 1.71\% increase in the Bj{\o}ntegaard-Delta bitrate (BD-rate). While VP9 provides several built-in speed levels which are designed to provide faster encoding at the expense of decreased rate-distortion performance, we find that our model is able to outperform the fastest recommended speed level of the reference VP9 encoder for the \textit{good} quality intra encoding configuration,  in terms of both speedup and BD-rate. 
\end{abstract}


% \fi
\IEEEpeerreviewmaketitle

\section{Introduction}
\label{sec:intro}
\IEEEPARstart{V}{p9} has been developed by Google \cite{vp9} as an alternative to mainstream video codecs such as H.264/AVC \cite{h264} and High Efficiency Video Coding (HEVC) \cite{hevc} standards. VP9 is supported in many web browsers and on Android devices, and is used by online video streaming service providers such as Netflix and YouTube.

\begin{figure}
\centering
\includegraphics[width=8cm]{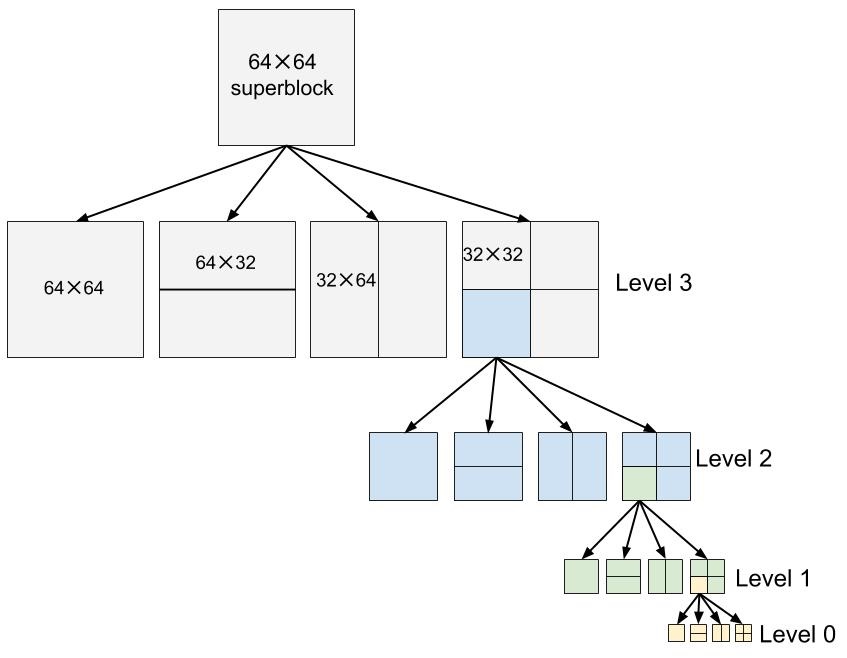}
%  \vspace{1.5cm}
\caption{\small{{Recursive partition of VP9 superblock at four levels showing the possible types of partition at each level.}}}
\label{fig:partition_types}
\end{figure}

As compared to both its predecessor, VP8 \cite{vp8} and H.264/AVC video codecs, VP9 allows larger prediction blocks, up to size $64 \times 64$, which results in a significant improvement in coding efficiency. In VP9, sizes of prediction blocks are decided by a recursive splitting of non-overlapping spatial units of size $64 \times 64$, called superblocks. This recursive partition takes place at four hierarchical levels, possibly down to $4 \times 4$ blocks, through a search over the possible partitions at each level, guided by a rate-distortion optimization (RDO) process. The Coding tree units (CTUs) in HEVC, which are analogous to VP9's superblocks, have the same default maximum size of $64 \times 64$ and minimum size of $8 \times 8$, which can be further split into smaller partitions ($4 \times 4$ in the intra-prediction case). However, while HEVC intra-prediction only supports partitioning a block into four square quadrants, VP9 intra-prediction also allows rectangular splits. Thus, there are four partition choices at each of the four levels of the VP9 partition tree for each block at that level: no split, horizontal split, vertical split and four-quadrant split. This results in a combinatorial complexity of the partition search space since the square partitions can be split further. A diagram of the recursive partition structure of VP9 is shown in Fig. \ref{fig:partition_types}.

Although the large search space of partitions in VP9 is instrumental to achieve its rate-distortion (RD) performance, it causes the RDO based  search to incur more computational overhead as compared to VP8 or H.264/AVC, making the encoding process slower. Newer video codecs, such as AV1 \cite{av1} and future Versatile Video Coding (VVC) \cite{vvc}, allow for prediction units of sizes from $128\times128$ to $4 \times 4$, giving rise to even deeper partition trees. As ultra high-definition (UHD) videos become more popular, the need for faster encoding algorithms will only escalate. One important way to approach this issue is to reduce the computational complexity of the RDO-based partition search in video coding. 

\IEEEpubidadjcol While state of the art performances in visual data processing technologies, such as computer vision, rely heavily on deep learning, mainstream video coding and compression technology remains dominated by traditional block-based hybrid codecs. However, deep learning based image compression techniques such as \cite{RNNimagecomp, GANcomp, variationalcomp} are also being actively explored, and have shown promise. A few of such image based deep learning techniques have also been extended to video compression with promising results  \cite{deepcoder, adversarialvideo, imageinterpolation}. A second category of work uses deep learning to enhance specific aspects of video coding, such as block prediction \cite{biprediction, interpred}, motion compensation \cite{fractionalinterpolation, motioncomp}, in-loop filtering \cite{inloop, residualinloop, multiscaleinloop}, and rate control \cite{ratecontrol}, with the objective of improving the efficiency of specific coding tools. Given these developments, there is a possibility of a future paradigm shift in the domain of video coding towards deep learning based techniques. Our present work is motivated by the success of deep learning-based techniques such as \cite{HEVCintra, inter&intra}, on the current and highly practical task of predicting the HEVC partition quad-trees. 

In this paper, we take a step in this direction by developing a method of predicting VP9 intra-mode superblock partitions in a novel bottom-up way, by employing a hierarchical fully convolutional network (H-FCN). Unlike previous methods of HEVC partition prediction, which recursively split blocks starting with the largest prediction units, our method predicts block merges recursively, starting with the smallest possible prediction units, which are $4 \times 4$ blocks in VP9. The bottom-up approach optimized using an  H-FCN model allows us to use a much smaller network than \cite{inter&intra}. By integrating the trained model with the reference VP9 encoder, we are able to substantially speed up intra-mode encoding at a reasonably low RD cost, as measured by the Bj{\o}ntegaard delta bitrate (BD-rate) \cite{bjontegaard}, which quantifies differences in bitrate at a fixed encoding quality level relative to another reference encode.  Our method also surpasses the higher speed levels of VP9 in terms of speedup, while maintaining a lower BD-rate as we show in the experimental results. 

\begin{figure*}
\centering
    \includegraphics[width=17cm]{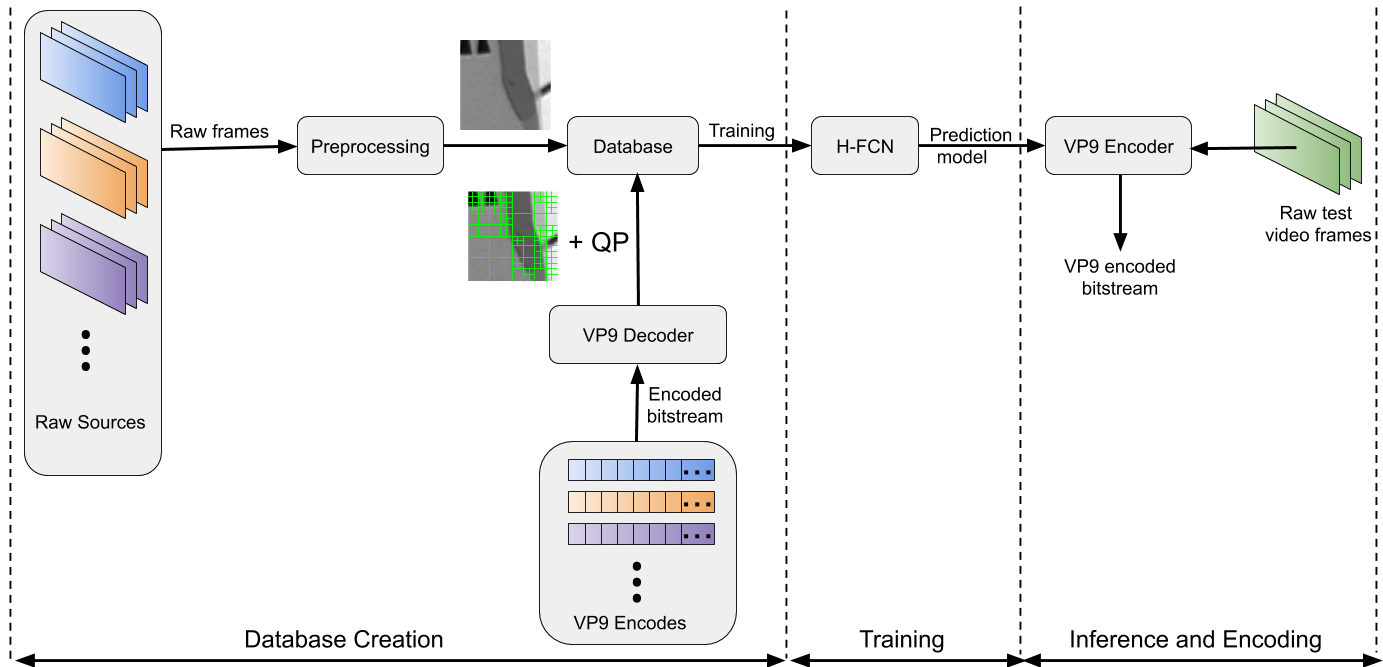}
    \caption{\small{{Flow diagram of our VP9 partition prediction. The inputs to the database are schematically shown for each superblock.}}}
    \label{fig:overview}
\end{figure*}

The main steps of our work presented in this paper can be summarized as follows:
\begin{enumerate}
\item We created a large and diverse database of VP9 intra encoded superblocks and their corresponding partition trees using video content from the Netflix library.
\item We developed a fast H-FCN model that efficiently predicts VP9 intra-mode superblock partition trees using a bottom-up approach. 
\item We integrated the trained H-FCN model with the VP9 encoder to demonstrably  speed up intra-mode encoding.  
\end{enumerate}
These steps are summarized by the block diagram in Fig. \ref{fig:overview}. The source code of our model implementation, including the modifications made to the reference VP9 decoder and encoder for creating the database of superblock partitions, and using the trained H-FCN model for faster intra encoding, respectively, is available online at \cite{repo}. 

The rest of the paper is organized as follows. In Section \ref{sec:relatedwork}, we briefly review earlier works relevant to the current task. Section \ref{sec:database} describes the VP9 partition database that we created to drive our deep learning approach. Section \ref{sec:method} elaborates the proposed method. Experimental results are presented in Section \ref{sec:results}. Finally, we draw conclusions and provide directions for future work in Section \ref{sec:conclusion}.

\section{Related Work}
\label{sec:relatedwork}
The earliest machine learning based methods that were designed to infer the block partition structures of coded videos from pixel data relied heavily on feature design. A decision tree based approach was used to predict HEVC partition quadtrees for intra frames from features derived from the first and second order block moments in \cite{icipML}, reportedly achieving a 28\% reduction in computational complexity along with 0.6\% increase in BD-rate. Using a support vector machine (SVM) classifier on features derived from measurements of the variance, color and gradient distributions of blocks, a 36.8\% complexity reduction was gained against a 3\% increase in BD-rate in \cite{screencontent}, on the screen content coding extension of HEVC in the intra-mode. 

With the advent of deep learning techniques in recent years, significant further breakthroughs were achieved \cite{intraCNN, inter&intra, HEVCintra}. In \cite{HEVCintra}, a parallel convolutional neural network (CNN) architecture was employed to reduce HEVC intra encoding time by 61.1\% at the expense of a 2.67\% increase in BD-rate. Three separate CNN models were used to learn the three-level intra-mode partition structure of HEVC in \cite{intraCNN}, obtaining an average savings of 62.2\% of encoding time against a BD-rate increase of 2.12\%. This approach was extended in \cite{inter&intra} to reduce the encoding time of both intra and inter modes using a combination of a CNN with a long short-term memory (LSTM) architecture. This approach reduced the average intra-mode encoding time by 56.9-66.5\%  against an increase of 2.25\% in BD-rate, while in the inter mode, a 43.8\%-62.9\% average reduction was obtained versus an increase of 1.50\% in  BD-rate. 

However, there has been little work reported on the related problem of reducing the computational complexity of RDO based superblock partition decisions in VP9, and even less  work employing machine learning techniques. A multi-level SVM based early termination scheme for VP9 block partitioning was adopted in \cite{mlvp9}, which reduced encoding time by 20-25\% against less than a 0.03\% increase in BD-rate in the inter mode. Although superblock partition decisions using RDO consume bulk of the compute expense of intra-mode encoding in VP9, to the best of our knowledge, there has been no prior work on predicting the complete partition trees of VP9 superblocks. 

The problem of VP9 superblock partition prediction is a hierarchical decision process, which involves choosing one of four types of partitions for each block, at every level of the partition tree. A hierarchical structure was introduced in \cite{iccvHDCNN}, using a two-level CNN that was trained to perform coarse-to-fine category classification, achieving improvements in the classification accuracy. Based on similar principles, \cite{bcnn} extended the hierarchical classification  approach to more than two levels, using a branched CNN model with multiple output layers yielding coarse-to-fine category predictions. The ability of these architectures to capture the hierarchical relationships inherent in visual data motivates our model design. At the same time, unlike global image tasks such as classification, inferring partition trees from superblock pixels is a spatially dense prediction task. For example, on a $64\times64$ superblock, there can be a maximum of 64 $8\times8$ blocks whose partitions are to be inferred. This means that as many as 64 localized predictions must be made on a $64 \times 64$ superblock at the lowest level of the partition tree. 

Fully convolutional networks (FCNs) have been shown to perform remarkably well on a variety of other spatially dense prediction tasks, such as semantic segmentation \cite{fcn, segnet}, depth estimation \cite{depth}, saliency detection \cite{saliency}, object detection \cite{objectdetection}, visual tracking \cite{visualtracking} and dense captioning \cite{densecap}. Moreover, convolution layers are typically faster than fully connected layers for similar input and output sizes. Likewise, we have found that the H-FCN model that we have developed and explain here imparts the ability to simultaneously handle hierarchical prediction and dense prediction with significant speedup. 

\section{VP9 Intra-mode Superblock Partition Database}
\label{sec:database}
In order to facilitate a data-driven training of our H-FCN model, we constructed a large database of VP9 intra encoded superblocks, corresponding QP values, and partition trees. In the absence of a publicly available superblock partition database for VP9 similar to the one developed in \cite{inter&intra} for HEVC CTU partitions, this was a necessary first step for our work. 

\subsection{Partition Tree Representation}
\label{subsec:outputrep}
\begin{figure}
\centering
\includegraphics[width=7.5cm]{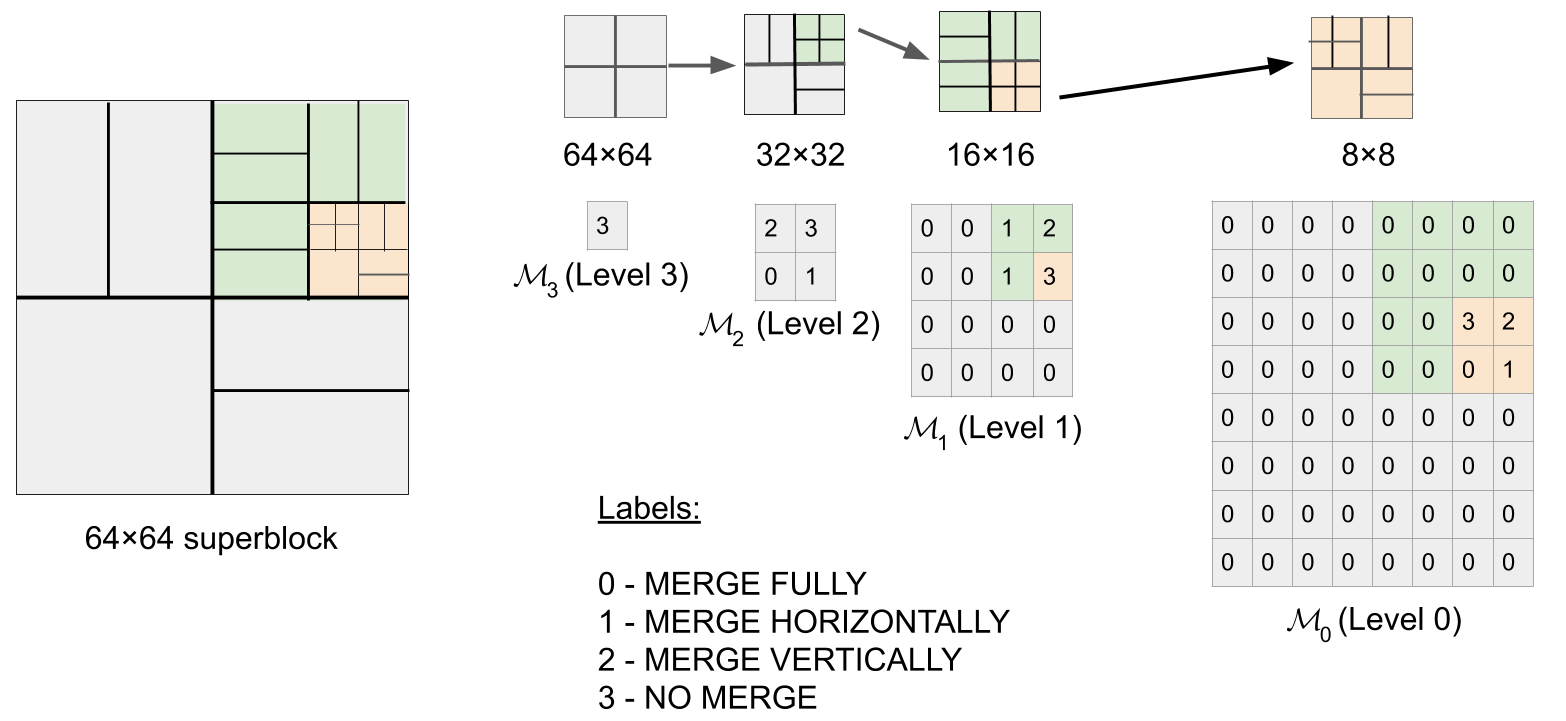}
%  \vspace{1.5cm}
\caption{\small{{Matrix representation of the four-level partition tree.}}}
\label{fig:partition_tree}
\end{figure}

Since the number of possible partition trees of a superblock is too large to be represented as distinct classes in a multi-class classification problem, we need a simple and concise description of the partition tree to ensure effective learning. The partition tree representation we adopt is similar to \cite{inter&intra}, but represents block merges instead of splits to facilitate bottom-up prediction. In our approach, the partition tree is represented by four matrices $\mathcal{M}_0, \cdots, \mathcal{M}_3$ that correspond to the four levels of the VP9 partition tree. An example of a superblock partition tree is illustrated in Fig. \ref{fig:partition_tree}. The four possible merges of the blocks at each level (including the possibility of no merge) are indicated by the numbers 0 to 3 as shown in Fig. \ref{fig:partition_tree}. Each element of the matrices indicates the type of merge of the group of four blocks corresponding to that element's location at that level. For example, each element of the $8\times8$ matrix $\mathcal{M}_0$ indicates how the four $4\times4$ blocks at corresponding locations are merged at level 0. Similarly, $\mathcal{M}_1$, $\mathcal{M}_2$ and $\mathcal{M}_3$ represent merges of non-overlapping groups of four $8\times8$, $16\times16$ and $32\times32$ blocks, respectively, at the higher levels. We denote the partition tree of a superblock by $\mathcal{P}=\{\mathcal{M}_0, \cdots, \mathcal{M}_3\}$. This succinct representation allows us to formulate the problem as a multi-level, multi-class classification task, with 4 levels, and 4 classes corresponding to each matrix element. 

\subsection{Database Creation}
\label{subsec:database}
Our database was created using content from the Netflix video catalog, encoded using the reference VP9 encoder from the \textit{libvpx} package \cite{libvpx}, in VP9 profile 0 (8 bits/sample and 4:2:0 chroma subsampling), using speed level 1 and the quality setting \textit{good}. The contents were drawn from 89 cinematic productions and 17 television episodes, each from a unique television series. The contents selected were drawn from different genres, such as action, drama, animation, etc. 
 
We encoded each content at three resolutions: $1920 \times 1080$, $1280 \times 720$  and $960 \times 540$. The partition pattern selected by the RDO on each superblock depends on both its visual content, as well as the QP value chosen for that superblock. Thus, we modified the VP9 decoder from the \textit{libvpx} package to record the computed partition tree $\mathcal{P}$ of each superblock, in the form described in Section \ref{subsec:outputrep}, along with the corresponding QP value $\mathcal{Q}$  while decoding the intra frames of the VP9 encoded bitstreams. 

In order to obtain the raw pixel data corresponding to the superblocks of each VP9 encoded video, the source videos were converted to a YCbCr 4:2:0 8-bit representation, then downsampled to the encode resolution via Lanczos resampling, if the source and the encode were at different resolutions. Following this process, the luma channels of non-overlapping $64\times64$ blocks were extracted from the source frames at the encode resolution. These steps constitute the preprocessing stage shown in Fig. \ref{fig:overview}. Denote this superblock pixel data by $\mathcal{S}$. Thus, each sample of our database may be expressed by a tuple $(\mathcal{S},\mathcal{Q},\mathcal{P})$. If the frame width or height (or both) is not exactly divisible by 64, the VP9 encoder zero pads the frame boundaries to construct superblocks of size $64 \times 64$ at the boundaries during encoding. However, we excluded the boundary superblocks with partial zero padding from our database. 

The HEVC intra-mode partition database \cite{inter&intra} is limited to only 4 QP values. By contrast, our database encompasses internal QP values in the range 8-105, where the internal QP value range for VP9 is 0-255 (the corresponding external QP value range is 0-63). The range of QP values used in our database is a practical range used for encoding intra frames in adaptive streaming, where rather than using higher QP values to stream videos at low target bitrates, a lower resolution video is streamed, which is suitably upsampled later at the viewers' end \cite{adaptivestreaming}. We refer the reader to Appendix B for a comparison of the encoding performance on our database against on the intra-mode database of \cite{inter&intra}. 

Our new database contains content from only a single episode of each television series. We divided our database into training and validation sets, as summarized in Table \ref{table:database_summary}, where the letters M and E indicate cinematic ``movies" and television ``episodes," respectively. The division was conducted such that there was no overlap in content between the training and validation sets, i.e. entirely different movies and television series were used for each set. Table \ref{table:database_summary} also specifies the percentage of computer graphics image (CGI) content in each set. Although the superblocks database of Netflix content cannot be made publicly available (due to the content licenses), we do provide the code for the modified VP9 decoder at \cite{repo}, which can be used to generate a similar database from a set of VP9 encoded bitstreams and their original source videos. The test set was created from publicly available content, different from the Netflix contents used in the training and validation sets, as described later in Section \ref{subsec:performance}.

%\begin{figure}
%\centering
%\includegraphics[width=7.5cm]{qp_histogram}
%%  \vspace{1.5cm}
%\caption{\small{{Empirical distribution of QP values of $10^5$ randomly drawn superblocks from the training set of our database.}}}
%\label{fig:QP_dist}
%\end{figure}

\begin{table}[htb]
\caption{Summary of VP9 intra-mode superblock partition database}
\centering
\vspace{0.1cm}
 \begin{tabular}{|c | c | c | c|} 
 \hline
 Database &  Contents & \% of CGI content & \# of samples \\ [0.5ex] 
 \hline
Training  & 62 (M) + 12 (E)  & 12.16 &  11,990,384  \\ 
 \hline
Validation & 27 (M) + 5 (E) & 12.50 & 4,698,195  \\
  \hline
\end{tabular}
\label{table:database_summary}
\end{table}

\section{Proposed Method}
\label{sec:method}
As mentioned earlier, in our approach the partition tree is constructed in a bottom-up manner, where merges of the smallest possible blocks, which are of size $4 \times 4$ in VP9, are predicted first at the lowest level of the partition tree, followed by merge predictions on larger blocks at the upper levels. This is unlike the coarse-to-fine approaches used in many image analysis applications, but it is well-motivated here. In a multi-layered CNN model, the early layers learn simple low level features depicting local image characteristics, such as edges and corners. These simple features are progressively  combined to form textures and more complex features representing higher level global semantics by the deeper layers. The intuition behind our bottom-up block merge strategy follows from this observation, which is supported by feature visualization techniques applied to successive CNN layers \cite{visualization}. Since the smaller blocks are more spatially localized, low-level local features, such as edges and corners that are learned by the early layers of a CNN, should be adequate to predict the merge types of $4\times4$ blocks i.e. the partition types of $8\times8$ blocks. The most notable benefit of this approach is that early prediction of the merge types of the smaller blocks at the lower levels of the partition tree, saves computation time and network parameters that can instead be invested in predicting the merge types of larger blocks at the upper levels, on which the CNN needs to learn more complex patterns. This allows us to design a deeper network for the higher levels, having many fewer trainable parameters than for example, the ETH-CNN of \cite{inter&intra} with three parallel branches and $1,287,189$ trainable parameters, which uses just three convolutional layers at each level of prediction, despite the larger number of trainable parameters. A deeper network is able to learn additional levels of hierarchical abstractions of the data, which is relevant in the context of hierarchical partition prediction in VP9. An analysis of the efficacy of this bottom-up prediction strategy is given in Appendix A, by comparing it with an alternative top-down strategy. 
\subsection{H-FCN Architecture}

\begin{figure*}
\centering
\includegraphics[width=17.5cm]{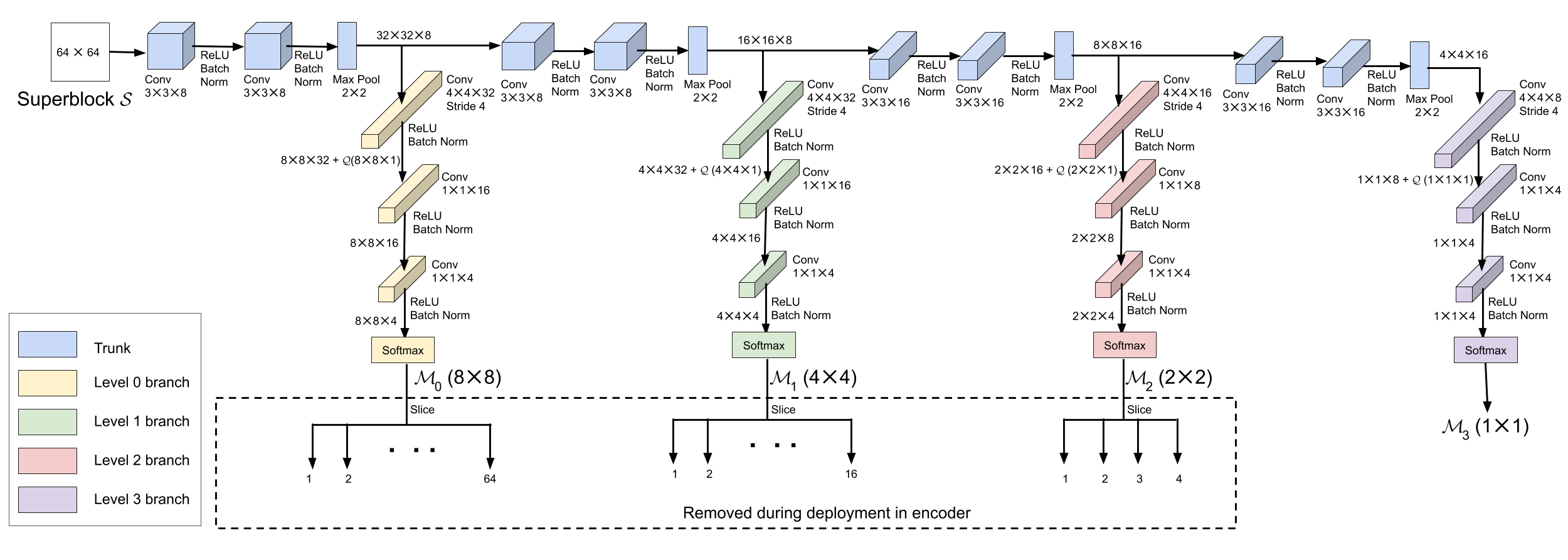}
%  \vspace{1.5cm}
\caption{\small{{H-FCN model architecture.}}}
\label{fig:HFCN}
\end{figure*}

The architecture of our H-FCN model is shown in Fig. \ref{fig:HFCN}. It consists of four output branches and a \textit{trunk} from which the branches emanate. This branched architecture is similar to \cite{bcnn}. However, unlike \cite{bcnn}, our model is fully convolutional and uses a bottom-up prediction scheme. The inputs to the model are the superblocks $\mathcal{S}$ and corresponding QP values $\mathcal{Q}$. The trunk has a conventional CNN structure with convolutional layers followed by rectified linear unit (ReLU) nonlinearities \cite{relu} and using batch normalization \cite{batchnorm}. Also $2\times2$ max pooling is applied following every two convolutional layers in the trunk. The outputs of the model are the matrices $\mathcal{M}_0, \cdots \mathcal{M}_3$ which constitute the partition tree $\mathcal{P}$. Each branch predicts one matrix from the lowest to the highest level of the partition tree, as shown in Fig. \ref{fig:HFCN}.  

The input to each branch are the features produced by the convolutional layers processed by ReLU, batch normalization and max pooling at each depth of the trunk. At the first layer of each branch, a convolutional layer having filters with spatial kernel dimensions of $4 \times 4$ and stride 4 is used, making it possible to isolate the features corresponding to adjacent blocks at that level. The QP value is fed at the input of the second convolutional layer of each branch by concatenating a  matrix of identical elements of value $\mathcal{Q}$ to the output of the first convolutional layer (after ReLU and batch normalization operations). The size of this matrix at each branch is chosen to match the spatial dimensions of the output of the first convolutional layer of that branch, which allows the two to be concatenated along their third dimension. Since the features that correspond to adjacent blocks are isolated by the first convolutional layer of each branch, feeding in QP value input in this manner also ensures that there is a copy of the QP value corresponding to the prediction of each block at the subsequent layers of that level. Unlike CNNs that make global predictions using fully connected output layers, we need to make structured local predictions at each branch, except for the last one (at the topmost level of the partition tree). The subsequent output layers of these branches are designed as convolutional layers having $1\times 1\times M$ filters, where $M$ is the ``depth" dimension of the input to the layers. This design serves to maintain  isolation between features that correspond to adjacent blocks by forming local connections to the outputs of the previous layers. At the last branch, the spatial dimension of the input to the first $1 \times 1$ convolutional layer is also $1\times 1$, which makes it functionally equivalent to a fully connected layer. Finally, the output of the last $1 \times 1$ convolutional layer of each branch is fed to a softmax function, yielding a set of class probabilities. 

Using $1\times 1\times M$ convolutions on  the features derived from each block at particular convolutional layer of each branch, is akin to having a fully connected layer (in the ``depth" dimension) for every block at the same level while sharing the weights. This design has two important consequences. First, it increases the inference and training speed as compared to using fully connected layers because of the greatly reduced number of connections. Second, it further speeds up convergence during training by reducing the number of trainable parameters and by ``augmenting" the data, since blocks of the same size but at different spatial locations within a superblock are used to train the convolution parameters.   

It should be noted that, although we experimented with larger model architectures, we found the architecture of Fig. \ref{fig:HFCN} to be best suited to the task, as elaborated later in Section \ref{sec:results}. We will refer to the model as shown in Fig. \ref{fig:HFCN} as the ``H-FCN model" in the rest of the paper, unless otherwise stated. 

\subsection{Loss Function}
Since there are four possible types of merges for each group of four blocks at each level, predicting the elements of the matrices $\mathcal{M}_0, \cdots, \mathcal{M}_3$ is a multi-class classification with four classes. By slicing each of the four matrices into their constituent elements, we obtain a total of 85 outputs. A categorical cross-entropy loss is then applied to each output:
\begin{equation}
L_q(\mathbf{w})= - \frac{1}{N}\sum_{i=1}^{N}\sum_{j=1}^{K}y_{i,j}\text{log}(p_{i,j}^{q}(\mathbf{w})) \  q=1, \cdots, 85
\end{equation}
where $N$ is the batch size, $K=4$ is the number of classes, $\mathbf{w}$ represents the weights of the network, and $p_{i,j}^{q}(\mathbf{w})$ is the softmax probability of the $i^{\text{th}}$ sample, predicted for the $j^{\text{th}}$ class at the $q^{\text{th}}$ output of the network. The net loss of the network is then $L(\mathbf{w})=\sum_{p=1}^{85}L_q(\mathbf{w})$.

\subsection{Integration with VP9 Encoder}\
\label{subsec:intg}
We integrated the trained H-FCN model with the reference VP9 encoder implementation, available in the \textit{libvpx} package \cite{libvpx}. 
By this integration, we replaced the RDO based partition search of the VP9 encoder with the H-FCN model prediction, which, as it turns out, makes intra encoding much faster.  While encoding each frame, the partition trees of all $64\times64$ superblocks completely contained within the frame boundaries were collectively predicted as a batch, using the integrated model. 

\begin{figure}
\centering
\includegraphics[width=7.5cm]{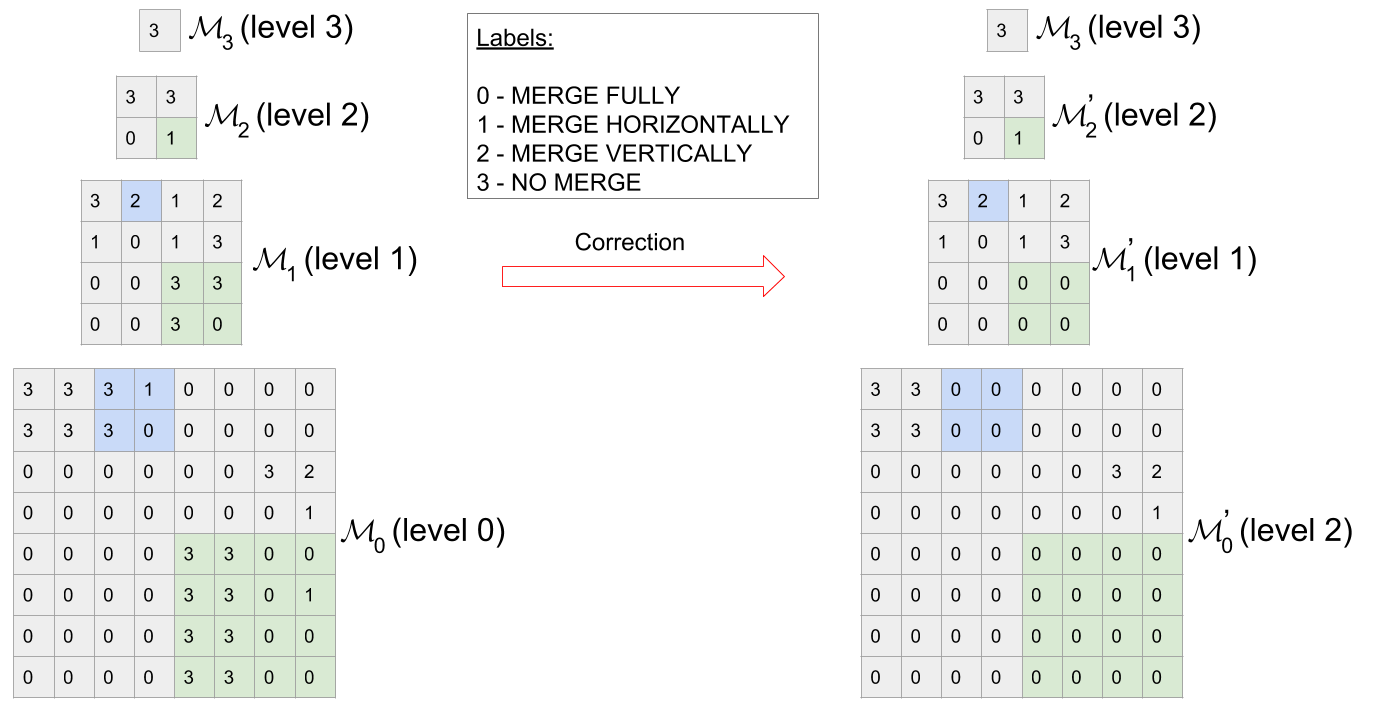}
%  \vspace{1.5cm}
\caption{\small{{Top-down inconsistency correction.}}}
\label{fig:correction}
\end{figure}

Since the different levels of the superblock partition are modeled independently of one another, the predictions of any two adjacent levels might be mutually inconsistent, producing an invalid partition tree. For example, a ``full merge" may be predicted for a group of four $16 \times 16$ blocks at level 2, whereas the four $8 \times 8$ subblocks within one of these $16 \times 16$ blocks can be predicted to have no merge in level 1. This is inconsistent, because in this case the merger of four $16 \times 16$ blocks indicates that a $32 \times 32$ block is not split, and thus the $16 \times 16$ subblocks within it cannot have a split either, although a split is required by the ``no merge" prediction of the corresponding group of $8 \times 8$ blocks. Thus, it may be necessary to correct the predictions in $\mathcal{P}$ to obtain a valid partition tree that can be used while encoding. We devised a top-down correction procedure that is illustrated in Fig. \ref{fig:correction}, where the colored regions are inconsistent  with predictions at other levels. Consistency is enforced between adjacent levels by correcting any of the other three merge predictions to ``full merge," at all such blocks at the lower of the two levels enclosed by a larger block predicted to have ``full merge" at the higher level. Beginning with level 2, predictions at each level are successively corrected to be consistent with its corrected next higher level in this manner. In other words, first $\mathcal{M}_2$ at level 2 is corrected to be consistent with $\mathcal{M}_3$; let the corrected matrix at level 2 be $\mathcal{M}^{'}_{2}$. At the next step, $\mathcal{M}_1$ is corrected to be consistent with $\mathcal{M}^{'}_{2}$ to obtain $\mathcal{M}^{'}_{1}$ and so on. At the end of this procedure, we have $\mathcal{P}^{'}=\{\mathcal{M}_3, \mathcal{M}^{'}_{2}, \mathcal{M}^{'}_{1}, \mathcal{M}^{'}_{0}\}$, where $\mathcal{P}^{'}$ is a valid partition tree. 

Although the predictions made by the H-FCN are independent at each level, they were found to be remarkably consistent across levels. Our experiments revealed that only about 5.17\% of the predicted superblock partition trees from the validation set were inconsistent, and thus needed the aforementioned correction. This suggests that although the consistency requirement was not explicitly enforced, the H-FCN model implicitly learns to make consistent merge predictions most of the time. The motivation behind our choice of the inconsistency correction approach is explained in Section \ref{subsec:performance}.

Predictions in $\mathcal{P}^{'}$  are then ordered to form a preorder traversal of the partition tree from left to right and top to bottom, which corresponds to the order in which the blocks are encoded in VP9. The predictions thus ordered are then recursively used to replace the RDO module of the encoder to decide the partitions of all the superblocks except those extending beyond frame boundaries. Since these boundary cases were not included in our database, we simply invoke the RDO module to encode them. 

\section{Experimental Results}
\label{sec:results}

In order to experimentally evaluate the performance of our H-FCN model relative to RDO, we compared the encoding time and RD performance of the two approaches on a set of test video sequences at three different resolutions. To further validate the efficacy of our approach, we also extended the comparison to include the highest recommended speed level  of VP9 for our encoding configuration (\textit{cpu-used} 4 at \textit{good} quality).

\subsection{System Settings}
We developed and trained our H-FCN model using Tensorflow (version 1.12) with the Keras API. The model was trained on a system with an Intel Core i7-6700K CPU @4 GHz, with 8 cores and 32 GB RAM running a 64 bit Ubuntu 16.04 operating system. The training was accelerated with a Nvidia Titan X Pascal GPU with 12 GB of memory. 

Since the loss is not calculated during inference, the matrix slicing operation to generate multiple outputs is unnecessary, and removing it reduces the inference time by a significant fraction in our Keras implementation. Thus, we removed the slicing layers during inference as indicated by the dashed box in Fig. \ref{fig:HFCN}, directly obtaining the matrices $\mathcal{M}_0, \cdots ,\mathcal{M}_3$ as the network outputs. The trained Keras model was then converted to a Tensorflow computation graph prior to deployment in the VP9 encoder. Integration of the trained model with the reference VP9 encoder was done using \textit{libvpx} version 1.6.0, which is same as the one used to encode the VP9 bitstreams to create the partition trees in our database. Since the \textit{libvpx} implementation of VP9 is in C language, we used the Tensorflow C API, which was compiled with support for optimizations that use Intel's Math Kernel Library for Deep Neural Networks. This enabled us to seamlessly embed our model within the VP9 encoder. The Tensorflow C API was also found to be considerably faster than both Keras and Tensorflow's native Python API for inference, which is an added advantage of this choice. While a GPU was used for training, all encoding tests were performed without a GPU on a single core Intel Core i7-7500U CPU @2.70GHz with 8 GB of RAM running  64 bit Ubuntu 16.04. 

\begin{figure}
\centering
\includegraphics[width=8.5cm]{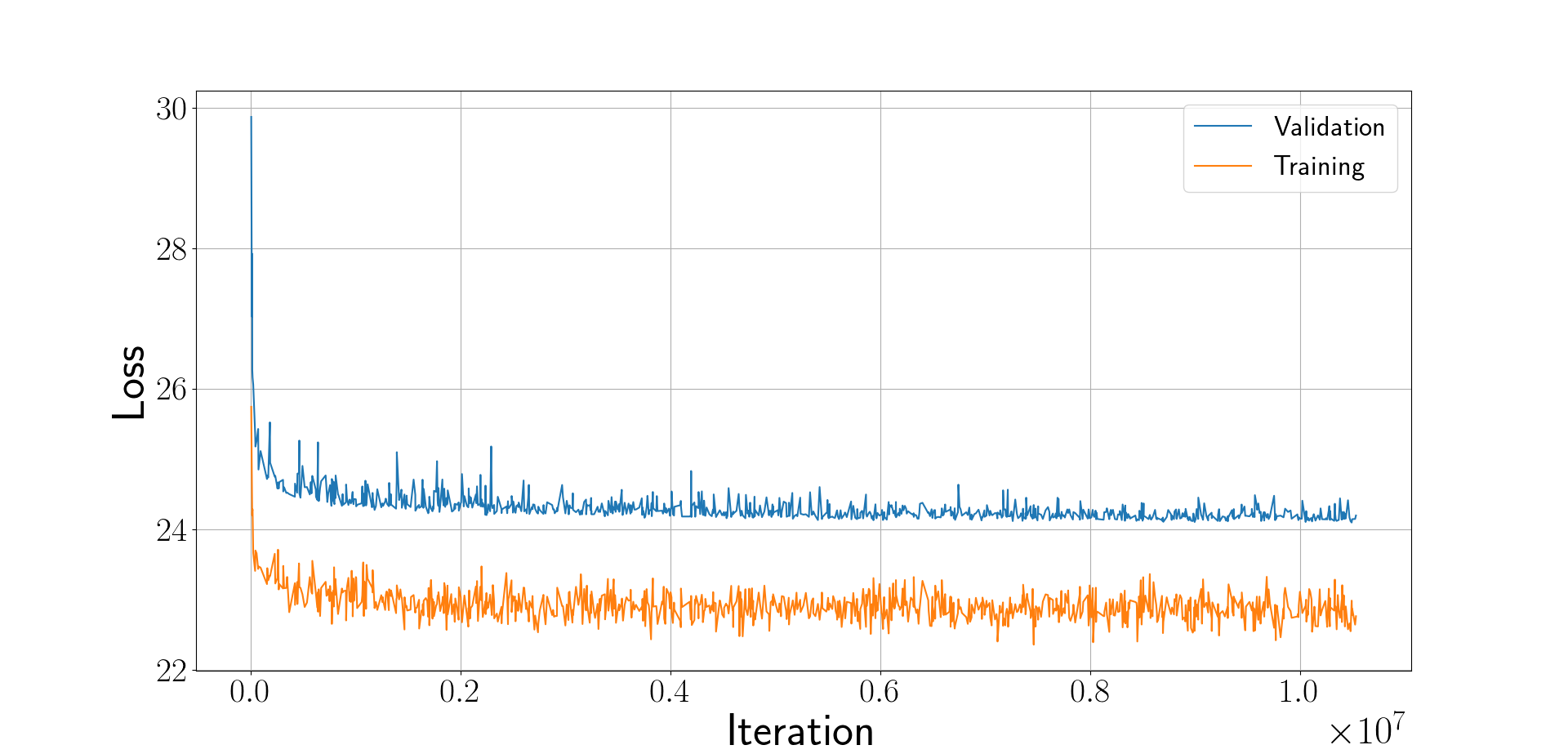}
%  \vspace{1.5cm}
\caption{\small{{H-FCN loss with training progress.}}}
\label{fig:loss}
\end{figure}

\begin{figure*} [htb]
    \centering
  \subfloat[QP=25\label{1a}]{%
       \includegraphics[width=0.49\linewidth]{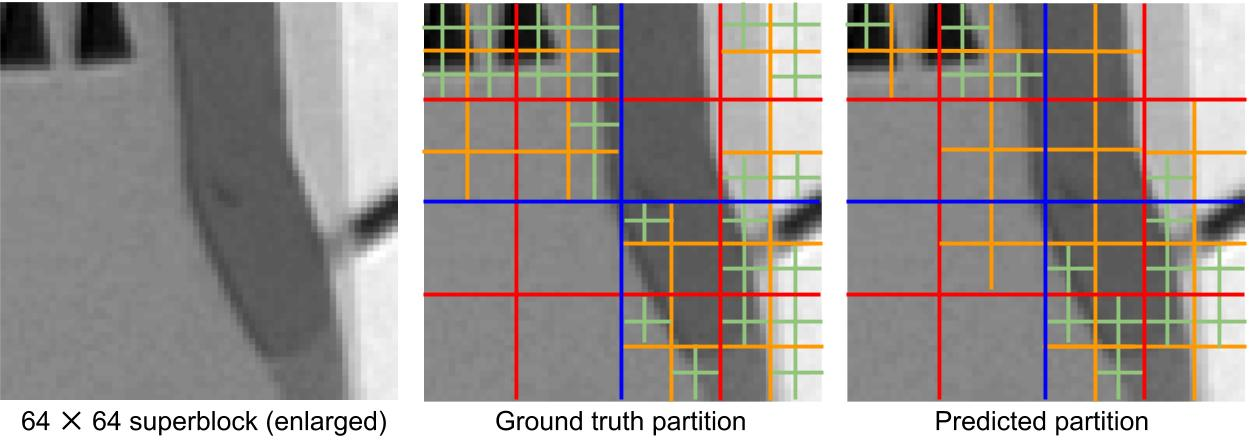}}
    \hfill
  \subfloat[QP=36\label{1c}]{%
        \includegraphics[width=0.49\linewidth]{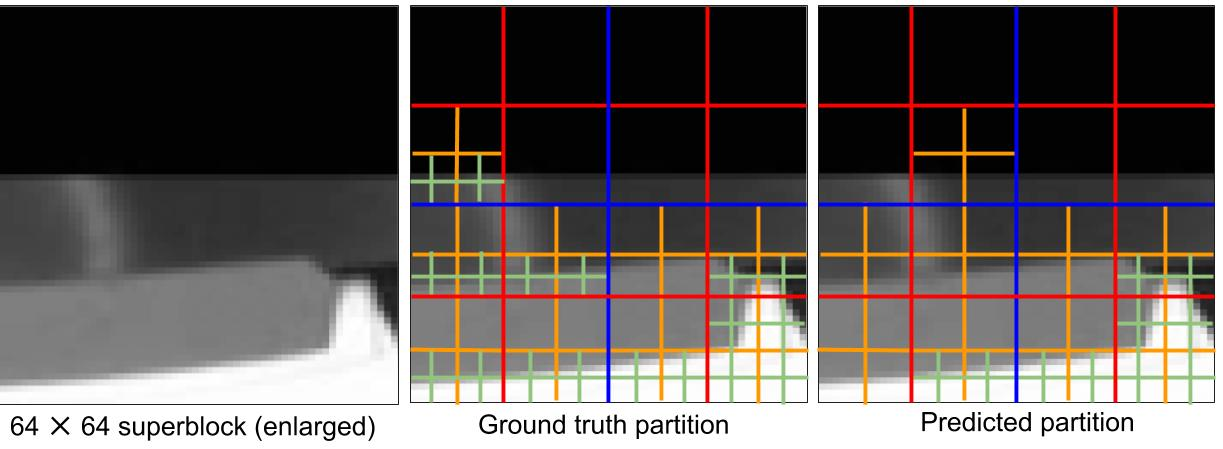}}
  \\
  \subfloat[QP=42\label{1b}]{%
   \includegraphics[width=0.49\linewidth]{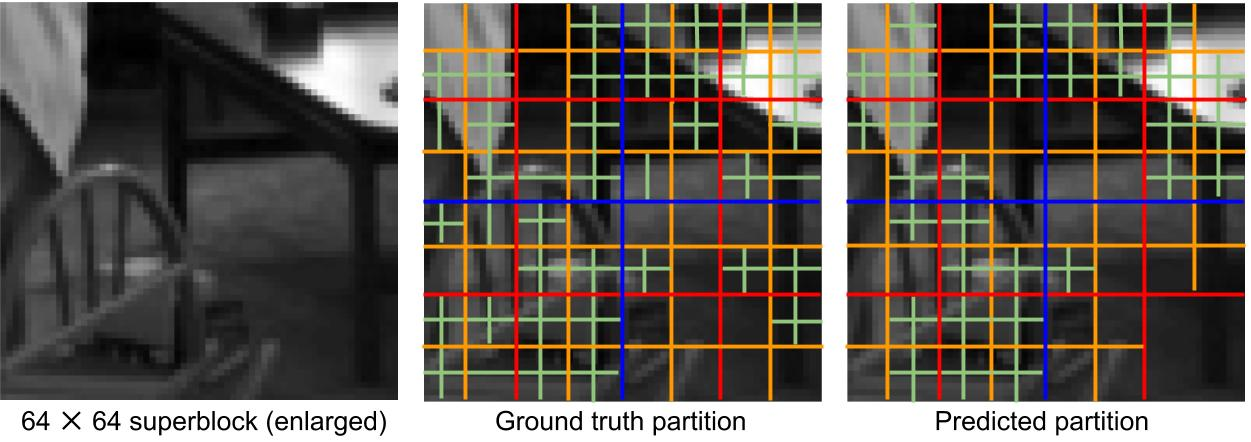}}
    \hfill
  \subfloat[QP=63\label{1d}]{%
        \includegraphics[width=0.49\linewidth]{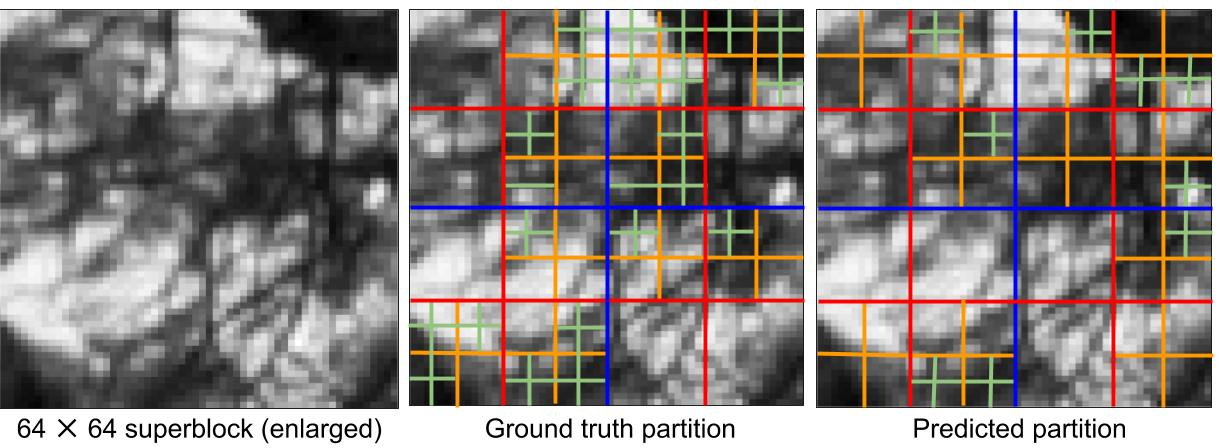}}
  \caption{Superblock partitions predicted by the trained H-FCN model compared with ground truth partitions on four superblocks encoded with different QP values.}
  \label{fig:visualization} 
\end{figure*}

\begin{table*}
  \caption{Prediction accuracy of H-FCN model}
  \centering
  \begin{tabular}{|p{2cm}|c|c|c|c|c|c|c|c|}
    \hline
    \multirow{2}{2cm}{\textbf{\# of parameters}} & \multicolumn{4}{|c|}{\textbf{Training (\%)}} & \multicolumn{4}{|c|}{\textbf{Validation (\%)}} \\
    % \hline
    \cline{2-9}
    & Level 0 & Level 1 & Level 2 & Level 3 & Level 0 & Level 1 & Level 2 & Level 3 \\
    %\hhline{~--}
    \hline
     \hfil 336,608 & 91.96 & 86.63 & 85.70 & 90.54 & 91.43 & 85.70 & 84.60 & 90.80 \\ \hline
    \hfil 70,408 & 91.89 & 86.46 & 85.33 & 90.23 & 91.34 & 85.53 & 84.28 &  90.81 \\ \hline
    \hfil 26,336 & 91.73 & 86.07 & 84.42 & 89.42 & 91.18 & 85.13 & 83.47 &  90.27 \\ \hline

  \end{tabular}
  \label{table:accuracy}
\end{table*}

\subsection{Training Details and Hyperparameters}
\label{subsec:hyperparams}
The H-FCN model as depicted in Fig. \ref{fig:HFCN} has $26,336$ trainable parameters and executes $\sim$10.8M floating point operations (FLOPs) per sample, at a per sample inference time of $\sim$1ms, where each sample in the inference batch is a $64 \times 64$ superblock.  We also experimented with two larger versions of the model, having $336,608$ and $70,408$ trainable parameters, which were identical to that in Fig. \ref{fig:HFCN}, except that we used a larger number of filters in certain layers.\footnote{The architectures of the  larger models are provided at \url{https://drive.google.com/file/d/1nDP42TjoeZhoWAlfhGlHlCMi04-CmKUE/view?usp=sharing}} The number of training and validation samples are as mentioned in Table \ref{table:database_summary}. The H-FCN model was trained with a batch size of 128 using the Adam optimizer \cite{adam},  with a step size of 0.001 for over $10^7$ iterations. The weights of each layer were initialized with randomly drawn samples from uniform distributions determined by the size of the inputs to each layer \cite{he}. Fig. \ref{fig:loss} illustrates the variation of the training and validation losses of the H-FCN model against the number of training iterations, where the validation loss was evaluated on $10^5$ samples from the validation set at the end of every 1000 iterations. 

\subsection{Prediction Performance}
\label{subsec:accuracy}

We evaluated the prediction accuracy of the trained model on the validation set. The average accuracies of the model at the four levels of prediction were evaluated on $10^5$ randomly drawn samples of the validation set, and are summarized in Table \ref{table:accuracy}. The corresponding performance on an equal number of random samples from the training set is also provided for reference. 

\begin{table*}[htb]
  \caption{Encoding performance of H-FCN model}
  \centering
  \begin{tabular}{|p{2.7cm}|p{2cm}|p{1.5cm}|p{1cm}|ccc|ccc|}
    \hline
    \multirow{2}{2.7cm}{\hfil{Sequence}} &  \multirow{2}{2cm}{Video Source} & \multirow{2}{1.5cm}{\hfil{Resolution}} & \multirow{2}{1cm}{\hfil{\# of frames}} & \multicolumn{3}{|c|}{$\Delta T$ (\%)} & \multicolumn{3}{|c|}{BD-rate (\%)} \\
    % \hline
    \cline{5-10}
    & &  &  & 336,608 & 70,408 & 26,336 & 336,608 & 70,408 & 26,336  \\
    %\hhline{~--}
    \hline
     \hfil{\textit{Sintel trailer}} & \hfil{Blender} & \hfil{\multirow{3}[20]{*}{1920$\times$1080}} &  \hfil{300} & -0.1 & 27.8 & \textbf{56.0}  & \textbf{2.78}  & 2.78  & 3.39 \\ 
     \hfil{\textit{Pedestrian area}} & \hfil{TUM} &  &  \hfil{375} & 36.6 & 41.6 & \textbf{61.8} & \textbf{2.08} & 2.12 & 2.45 \\ 
     \hfil{\textit{Sunflower}} & \hfil{TUM} &   &  \hfil{500} & 18.8 & 34.4 & \textbf{55.0} & \textbf{2.61} & 2.76 & 3.07 \\ 
     \hfil{\textit{Crowd run}} & \hfil{VQEG} &  &   \hfil{500} & 45.2 & 49.7 & \textbf{65.3} & \textbf{0.99} & 1.13 & 1.16 \\
     \hfil{\textit{Ducks takeoff}} & \hfil{VQEG} &  & \hfil{500} & 43.6 & 48.6 & \textbf{ 67.9} & \textbf{1.42} & 1.55  & 1.72 \\ 
     \hfil{\textit{Narrator}} & \hfil{Netflix El fuente} &  & \hfil{300} & 36.6 & 55.8 & \textbf{73.3} & 0.32 & \textbf{0.25}  & 0.58 \\
     \hfil{\textit{Food market}} & \hfil{Netflix El fuente} &  &  \hfil{300} & 37.0 & 46.4 & \textbf{68.5} & 1.51 & \textbf{1.50 } & 1.63 \\ 
     \hfil{\textit{Toddler fountain}} & \hfil{Netflix Chimera} &  &  \hfil{420} & 47.8 & 48.9 & \textbf{70.2} & 1.37 & 1.35  & \textbf{1.33} \\
     \hfil{\textit{Rainroses}} & \hfil{EBU} &  &  \hfil{500} & 57.9 & 64.2 & \textbf{73.6} & \textbf{0.83} & 0.86  & 0.92 \\ 
     \hfil{\textit{Kidssoccer}} & \hfil{EBU} &  &  \hfil{500} & 74.6 & 76.5 & \textbf{83.5} & 0.65 & \textbf{0.59} & 0.74 \\ 
     \hfil{Average} & \hfil{-} &  &  \hfil{-} & 39.8 & 49.4 & \textbf{67.5} & \textbf{1.46} & 1.49 & 1.70 \\ \hline
     
     \hfil{\textit{Big buck bunny}} & \hfil{Blender} & \hfil{\multirow{3}[20]{*}{1280$\times$720}} &  \hfil{300} & 37.6 & 55.2 & \textbf{69.3} & 2.26 & \textbf{2.22} & 3.21 \\ 
     \hfil{\textit{Park run}} & \hfil{TUM} &  &  \hfil{504} & 24.0 & 50.4 & \textbf{65.2} & \textbf{1.29} & 2.06 & 1.74 \\ 
     \hfil{\textit{Shields}} & \hfil{TUM} &  &  \hfil{504} & 54.2 & 62.9 & \textbf{69.3}& \textbf{1.61} & 1.88 & 1.73 \\
     \hfil{\textit{Into tree}} & \hfil{VQEG} &  &  \hfil{500} & 52.3 & 69.4 & \textbf{77.0} & \textbf{1.10} & 1.23  & 1.27 \\ 
     \hfil{\textit{Old town cross}} & \hfil{VQEG} &  &  \hfil{500} & 59.4 & 67.8 & \textbf{77.5} & \textbf{0.38} & 0.71  & 0.82 \\
     \hfil{\textit{Crosswalk}} & \hfil{Netflix El fuente} &  &  \hfil{300} & 42.4 & 63.3 & \textbf{74.4} & \textbf{0.91 }& 1.07  & 1.32 \\ 
     \hfil{\textit{Tango}} & \hfil{Netflix El fuente} &  &   \hfil{294} & 35.8 & 49.6 & \textbf{75.4} & \textbf{1.52} & 1.61  & 1.73 \\
     \hfil{\textit{Driving POV}} & \hfil{Netflix Chimera} &  &  \hfil{500} & 35.6 & 55.9 & \textbf{70.1} & 1.25 & \textbf{1.24}  & 1.46 \\ 
     \hfil{\textit{Dancers}} & \hfil{Netflix Chimera} &  &  \hfil{500} & 25.5 & 49.9 & \textbf{71.9} & 3.08 & \textbf{1.68}  & 2.47 \\ 
     \hfil{\textit{Vegicandle}} & \hfil{EBU} &  &  \hfil{500} & 47.1 & 54.0 & \textbf{71.6} & \textbf{1.47} & 1.61  & 1.74 \\ 
     \hfil{Average} & \hfil{-} &  & \hfil{-} & 41.4 & 57.8 & \textbf{72.2} & \textbf{1.49} & 1.53  & 1.75  \\ \hline

      \hfil{\textit{Elephant's dream}} & \hfil{Blender} & \hfil{\multirow{3}[20]{*}{960$\times$540}} & \hfil{545}   & 37.6 & 54.7 & \textbf{59.9} & \textbf{1.68} & 1.72 & 1.82 \\ 
     \hfil{\textit{Euro Truck Simulator 2}} & \hfil{Twitch} &  & \hfil{300}  & 48.0 & 68.0 & \textbf{73.1} & 1.65 & \textbf{1.63} & 1.69 \\ 
     \hfil{\textit{Station}} & \hfil{TUM} &  & \hfil{313}  & 64.0 & 65.4 & \textbf{80.7} & \textbf{2.01} & 2.08 & 2.27 \\
     \hfil{\textit{Rush hour}} & \hfil{TUM} &  &  \hfil{500}  & 27.3 & 50.3 & \textbf{66.4} & \textbf{1.71} & 1.92  & 1.98 \\ 
     \hfil{\textit{Touchdown pass}} & \hfil{VQEG} &  & \hfil{570}  & 45.2 & 63.7 & \textbf{72.8} & \textbf{1.45} & 1.49  & 1.62\\
     \hfil{\textit{Snow mnt}} & \hfil{VQEG} &  &  \hfil{570}  & 35.1 & 50.8 & \textbf{64.6} & 1.09 & 1.05  & \textbf{1.04} \\ 
     \hfil{\textit{Roller coaster}} & \hfil{Netflix El fuente} &  & \hfil{500}  & 46.1 & 53.6 & \textbf{73.9} & \textbf{1.40} & 1.46  & 1.68 \\
     \hfil{\textit{Dinner scene}} & \hfil{Netflix Chimera} &  &  \hfil{500}  & 23.9 & 54.0 & \textbf{65.7} & 1.48 & \textbf{1.38}  & 1.70 \\ 
     \hfil{\textit{Boxing practice}} & \hfil{Netflix Chimera} &  &  \hfil{254} & 43.1 & 58.4 & \textbf{68.6} & 1.43 & \textbf{1.42} & 1.50 \\ 
     \hfil{\textit{Meridian}} & \hfil{Netflix Meridian} & & \hfil{500}  &  42.4 & 57.9 & \textbf{69.6} & \textbf{1.36} & 1.46 & 1.47   \\
     \hfil{Average} & \hfil{-} &  & \hfil{-} & 41.3 & 57.7 & \textbf{69.5} & \textbf{1.53} & 1.56  & 1.68  \\ \hline
      \multicolumn{4}{|c|}{Overall average}  & 40.8 & 55.0 & \textbf{69.7} & \textbf{1.49} & 1.53 & 1.71 \\ \hline

  \end{tabular}
  \label{table:performance}
\end{table*}

In Table \ref{table:accuracy}, we also include the performance results of the two larger models with $336,608$ and $70,408$ parameters respectively, to show that our model design allows for smaller architectures with no significant impairment of prediction accuracy. From Table \ref{table:accuracy}, we observe that the maximum decline in accuracy on the validation set between the largest and the smallest models was 1.13\%, which occured at level 2. The impact of this decline in accuracy on the BD-rate is examined in Section \ref{subsec:performance}.

The partitions predicted by the H-FCN model are visualized in Fig. \ref{fig:visualization}, against the corresponding ground truth partitions, on four randomly selected superblocks encoded at different QP values from the validation set,  with the $64 \times 64$ superblocks shown enlarged for visual clarity. These visualizations serve to demonstrate that the global structure and density of the superblock partitions predicted by our H-FCN model are in good agreement with ground truth, although some local blocks may be partitioned differently between the two.

\subsection{Encoding Performance}
\label{subsec:performance}
Although the encoding time can be reduced by predicting the block partitions using a trained model instead of conducting RDO-based exhaustive search, the RD performance of learned models may suffer due to incorrect partition predictions. Thus, to evaluate the performance of our trained model, it is also necessary to assess its RD performance with respect to that of the reference RDO. Accordingly, we encoded several test video sequences with the original VP9 encoder using the RDO based partition search, and also with the modified VP9 encoder using the integrated H-FCN model to conduct partition prediction. The test video sequences were obtained as raw videos from multiple publicly available sources commonly used for evaluating video codecs.\footnote{Our test video sequences were sourced from \url{https://media.xiph.org/video/derf/} and \url{https://tech.ebu.ch/hdtv/hdtv_test-sequences}.} All of  the test sequences were converted to the YCbCr 4:2:0 8-bit format prior to encoding. The lengths of the sequences varied between 254 to 570 frames. Sequences which were originally longer were clipped to less than 550 frames (the number of frames used for each test sequence are reported in Table \ref{table:performance}). As mentioned earlier in Section \ref{subsec:database}, our model was jointly trained on three different spatial resolutions.  For each resolution, the test sequences chosen were at the same or higher resolution. The sequences that were originally at higher resolutions were downsampled to the required resolution using Lanczos resampling. 

A set of five internal QP values given by $\{15, 31, 47, 70, 99\}$ (corresponding to external QP values of $\{20, 30, 35, 40, 45\}$ respectively) was selected to represent a practical quality range of intra-frames used in adaptive streaming.  Each sequence was then encoded  at these five QP values in one pass, using constant quality intra-mode, one tile per frame, speed level 1 and the \textit{good} quality setting. These settings were chosen to be compatible with the encoding configuration used for our database, as mentioned in Section \ref{subsec:database}, although our model is equally amenable to be trained for other configurations and can be parallelized over multiple tiles. The RDO based VP9 encoder at these settings thus formed the baseline for our approach. We used the same encoding settings as the baseline for the modified encoder with the integrated H-FCN model. 

The percentage speedup of our method with respect to the RDO baseline for a test sequence is calculated as:
\begin{equation}
\Delta T = \frac{T_\text{RDO}-T_\text{H-FCN}}{T_\text{RDO}} \times 100
\label{eq:deltaT}
\end{equation}
where $T_\text{RDO}$ and $T_\text{H-FCN}$ are the total times taken to encode the sequence at all  five QP values with the RDO baseline encoder and the H-FCN integrated encoder, respectively. Since our model is integrated with the encoder, $T_\text{H-FCN}$ includes the total inference time of the H-FCN model on the superblock batches from all of the frames of the video sequence to be encoded, as well as the encoding time without the RDO procedures for superblock partition decision. Thus, a positive value of $\Delta T$ represents a speedup with respect to the RDO. To measure the RD performance, we also computed the BD-rate relative to the RDO baseline, using peak signal-to-noise ratio (PSNR) as the distortion metric. Table \ref{table:performance} reports the $\Delta T$ and BD-rate values of the test sequences when using the H-FCN models  with $336,608$, $70,408$ and $26,336$ parameters. The test videos in Table \ref{table:performance} comprise 30 distinct contents, equally divided between the three resolutions. In Table \ref{table:performance}, the first sequences of the $1920\times 1080$ and  $1280\times720$ categories as well as the first two sequences of the $540 \times 960$ category are CGI content. As would be expected, the smallest model, with just $26,336$ parameters, achieved the highest speedup over all the test videos. Although the model with $336,608$ parameters achieved the highest accuracy on the validation set, we observe that for some test sequences, the smaller models achieved lower BD-rate values. The overall BD-rates achieved by the three models were 1.49\%, 1.53\% and 1.71\%. The smallest model was thus found to maintain a reasonable BD-rate, while achieving the most speedup. All of the analysis that we provide was conducted using the smallest H-FCN model having $26,336$ parameters, unless otherwise mentioned. 
\begin{figure*}[htb]
\centering
\includegraphics[width=18cm]{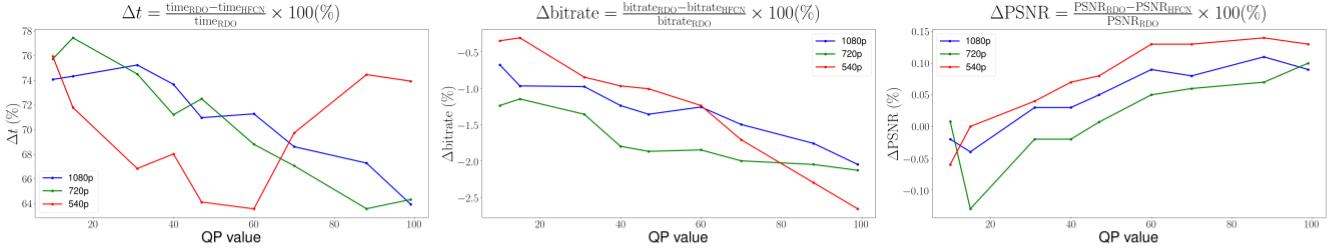}
%  \vspace{1.5cm}
\caption{\small{{Variation of $\Delta t$, $\Delta \text{bitrate}$ and $\Delta \text{PSNR}$ across the range of QP values at different resolutions.}}}
\label{fig:variation}
\end{figure*}

In order to show that the H-FCN model is able to generalize irrespective of the QP values chosen for evaluation within the range of the database, we also report the performance measures for an additional set of four internal QP values $\{10, 40, 60, 88\}$ in Table \ref{table:extraqp}. The speedup and BD-rate values reported in Table \ref{table:extraqp} are close to the corresponding values from Table \ref{table:performance}, indicating that the performance is consistent across the range of QP values used in the database. For each of the nine QP values in the combined set $\{10,15, 31, 40, 47, 60, 70, 88, 99\}$, the average changes in encoding time, bitrate and PSNR with respect to the RDO baseline, denoted by $\Delta t$, $\Delta \text{bitrate}$ and $\Delta \text{PSNR}$ respectively, are also plotted in Fig. \ref{fig:variation} for the three resolutions. The curves in Fig. \ref{fig:variation}, illustrate that a substantial speedup is obtained at every QP value, for a relatively small bitrate increase and PSNR reduction with respect to the RDO baseline. Although the encoding time was predominantly reduced with increases in the QP value using the RDO baseline as well as the proposed method, on the 540p sequences, the average encoding time of the RDO baseline was found to increase at the higher end of the range of QP values, while the proposed method was  still faster at these QPs. Consequently, the average $\Delta t$ values for the 540p sequences increased at the higher end of the QP range, unlike the trend observed for the other two resolutions.  Further, the $\Delta t$, $\Delta \text{bitrate}$ and $\Delta \text{PSNR}$ values obtained at the adjacent QP values tested are generally quite close. Subsequent results are hence reported using the initial set of five QP values. 

%\begin{equation}
%\begin{split}
%\Delta t &= \frac{t_\text{RDO}-t_\text{H-FCN}}{t_\text{RDO}} \times 100 \\
%\Delta \text{bitrate} &= \frac{\text{bitrate}_\text{RDO}-\text{bitrate}_\text{H-FCN}}{t_\text{RDO}} \times 100 \\
%\Delta \text{PSNR} &= \frac{\text{PSNR}_\text{RDO}-\text{PSNR}_\text{H-FCN}}{t_\text{RDO}} \times 100 
%\end{split}
%\end{equation}

\begin{table}[htb]
  \caption{Encoding performance on additional QP values}
  \centering
  \begin{tabular}{|p{1.2cm}|c|c|}
  \hline
     Resolution & $\Delta T$ (\%) & BD-rate (\%) \\
        \hline
   \hfil{1080p}  & 69.0 & 1.80  \\ \hline
   \hfil{720p}  & 67.8 & 1.37 \\ \hline
   \hfil{540p}  & 71.9 &  1.41  \\ \hline
   \hfil{Overall}  & 70.7  & 1.61 \\ \hline
     \end{tabular}
     \label{table:extraqp}
 \end{table}
 
%Continuing our informal comparison with \cite{inter&intra} in Section \ref{subsec:accuracy}, it is interesting to note that our approach achieves a higher speedup at a lower BD-rate for VP9, than \cite{inter&intra} does for HEVC in intra-mode. When the method of \cite{inter&intra} was applied to encode video sequences in intra-mode, an average speedup of 61.8\% was reported over four QP values (22, 27, 32 and 37) yielding an aggregate BD-rate  increase of 2.25\%,  as compared to the overall average speedup 69.7\% achieved by our method with an increase of 1.71\%  in BD-rate, when encoded at the external QPs values mentioned earlier.  It should be noted that the range of quantization levels that can be set while encoding is 0-51 for HEVC whereas it is 0-63 for VP9. Thus,  although the quantization levels chosen in the two cases are somewhat different, with \cite{inter&intra} being evaluated at higher quantization levels, we believe that the range of QP values chosen in our experiments is better suited for adaptive streaming, where low target bitrates are achieved by streaming at lower resolutions instead of using higher QP values \cite{adaptivestreaming}. 

We also evaluated the performance trade-off achieved by the top-down inconsistency correction scheme described in Section \ref{subsec:intg}, by exploring an alternative approach of resolving inconsistencies, whereby the RDO was invoked to decide the partitioning of a superblock if its predicted partition tree was determined to be inconsistent. Among all approaches that could be devised to handle inconsistencies, assigning superblocks having inconsistent partition predictions to the RDO in this manner would ensure the best performance in terms of BD-rate, albeit at the possible expense of a decline in the speedup achieved. This trade-off is summarized in Table \ref{table:inconsistency}, in terms of  average BD-rate and speedup of our inconsistency correction approach against that obtained when the RDO is employed to resolve inconsistencies, for each of the three resolutions considered in our work. Thus, by using the RDO to handle inconsistencies, an overall improvement of 0.4\% in BD-rate was achieved at the expense of a 6.3\% reduction in speedup on the sequences tested. Naturally, the improvement in RD performance that can be achieved by any other correction scheme is upper bounded by 0.4\%, and more sophisticated correction approaches such as those based on prediction probabilities indeed achieved negligible gains. Thus, due to its simplicity, we found that the top-down approach described  in Section \ref{subsec:intg}  to be a good choice for inconsistency correction. 

\begin{table}[htb]
  \caption{Performance trade-off of the inconsistency correction approach with respect to using RDO to resolve inconsistencies}
  \centering
  \begin{tabular}{|p{1.2cm}|cc|cc|}
    \hline
     \multirow{2}{1.5cm}{\hfil{Resolution}} & \multicolumn{2}{|c|}{$\Delta T$ (\%)} & \multicolumn{2}{|c|}{BD-rate (\%)} \\
    % \hline
    \cline{2-5}
     & Top-down & RDO & Top-down & RDO   \\
    \hline
   \hfil{1080p}  & 67.5 & 63.7 & 1.70  & 1.40  \\ \hline
   \hfil{720p}  & 72.2 & 62.3 & 1.75  & 1.13 \\ \hline
   \hfil{540p}  & 69.5 & 64.1 & 1.68  & 1.40  \\ \hline
   \hfil{Overall}  & 69.7 & 63.4 & 1.71  & 1.31  \\ \hline
     \end{tabular}
     \label{table:inconsistency}
 \end{table}

\begin{table*}[htb]
  \caption{Comparison of speedup versus BD-rate tradeoff of our approach with VP9 speed level 4}
  \centering
  \begin{tabular}{|p{2.7cm}|p{0.8cm}|cc|cc|cc|}
    \hline
    \multirow{2}{2.7cm}{\textbf{\hfil{Sequence}}} & \multirow{2}{0.8cm}{\textbf{\hfil{Res.}}} & \multicolumn{2}{|c|}{$\Delta T$ (\%)} & \multicolumn{2}{|c|}{BD-rate (\%)} & \multicolumn{2}{|c|}{BD-PSNR (dB)}\\
    % \hline
    \cline{3-8}
    &    & speed 4 & H-FCN & speed 4 & H-FCN & speed 4 & H-FCN \\
    %\hhline{~--}
    \hline
    \hfil{\textit{Sintel trailer}}  & \hfil{\multirow{3}[20]{*}{1080p}}  & 46.3  & \textbf{56.0}  & 4.46  & \textbf{3.39} & -0.21 &\textbf{-0.17}  \\ 

    \hfil{\textit{Pedestrian area}}  &   & \textbf{62.8} & 61.8 & 2.94 & \textbf{2.45}  & -0.12 & \textbf{-0.10}\\ 
    \hfil{\textit{Sunflower}}  &   & 50.2 & \textbf{55.0} & 6.74 & \textbf{3.07} & -0.28 & \textbf{-0.13} \\ 
     \hfil{\textit{Crowd run}} &  & 62.8 & \textbf{65.3}  & 2.30 & \textbf{1.16} & -0.21 & \textbf{-0.11} \\ 
     \hfil{\textit{Ducks takeoff}} & & 63.2 & \textbf{67.9} & 1.88  & \textbf{1.72} & -0.18 &\textbf{ -0.17}  \\ 
     \hfil{\textit{Narrator}}  &  & 59.5 & \textbf{73.3} & 3.72 & \textbf{0.58} & -0.13 & \textbf{-0.02} \\
     \hfil{\textit{Food market}} &  & 55.9 & \textbf{68.5} & 2.39 & \textbf{1.63}  & -0.18 & \textbf{-0.12} \\ 
     \hfil{\textit{Toddler fountain}}  &  & 61.7 & \textbf{70.2} & 1.73 & \textbf{1.33}  & -0.14 & \textbf{-0.11}\\
     \hfil{\textit{Rainroses}}  &   & 72.9  & \textbf{73.5}  & 2.50 & \textbf{0.92}  & -0.14 & \textbf{-0.04} \\ 
     \hfil{\textit{Kidssoccer}}   &   & \textbf{85.1} & 83.5 & 0.86 & \textbf{0.74} & -0.10 & \textbf{-0.08} \\ 
     \hfil{Average}   &   & 62.0 & \textbf{67.5} & 2.95 & \textbf{1.70} & -0.17 & \textbf{-0.10} \\  \hline
     \hfil{\textit{Big buck bunny}}  &  \hfil{\multirow{3}[20]{*}{720p}} & 58.4 & \textbf{69.3} & 7.92  & \textbf{3.21} & -0.51 & \textbf{-0.21}  \\ 
    \hfil{\textit{Parkrun}}  &  & 61.1  & \textbf{65.1} & 3.63  & \textbf{1.74} & -0.46 & \textbf{-0.22} \\ 
     \hfil{\textit{Shields}} & & \textbf{70.4} & 69.3 & 2.84 &  \textbf{1.73} & -0.24  & \textbf{-0.14} \\ 
     \hfil{\textit{Into tree}}  &  & 75.6  & \textbf{77.0} & 1.57 & \textbf{1.27} & -0.13 & \textbf{-0.11} \\ 
     \hfil{\textit{Old town cross}}  & & 75.5 & \textbf{77.5} & 2.00 &  \textbf{0.82} & -0.15 & \textbf{-0.06} \\ 
     \hfil{\textit{Crosswalk}} &  & 71.8  & \textbf{74.4} & 3.75 & \textbf{1.32}  & -0.15 & \textbf{-0.05}\\ 
     \hfil{\textit{Tango}} &  & 62.4  & \textbf{75.3} & 2.39  & \textbf{1.73}  & -0.11 & \textbf{ -0.08} \\ 
     \hfil{\textit{Driving POV}} & & 62.6 & \textbf{70.1} & 2.00 &   \textbf{1.46} &  -0.16 & \textbf{-0.12} \\ 
     \hfil{\textit{Dancers}}  &  & 70.4  & \textbf{71.9} & 12.26  &  \textbf{2.47} & -0.09 & \textbf{ -0.00} \\ 
     \hfil{\textit{Vegicandle}} &  & \textbf{74.1} & 71.6 & 2.85  & \textbf{1.74}  & -0.14 & \textbf{-0.08}  \\ 
     \hfil{Average}   &   & 68.2 & \textbf{72.2}& 4.12 & \textbf{1.75} & -0.21 & \textbf{-0.11} \\  \hline
     \hfil{\textit{Elephants dream}}  &  \hfil{\multirow{3}[20]{*}{540p}} & 59.2  & \textbf{59.9} & 2.13 & \textbf{1.82} &  -0.18 & \textbf{-0.16} \\ 
    \hfil{\textit{Euro Truck Simulator 2}} & & 72.0 & \textbf{73.1} & 1.89  & \textbf{1.69}  & -0.23 & \textbf{-0.20} \\ 
     \hfil{\textit{Station}}  &  & 76.5 & \textbf{80.7} & \textbf{2.06} & 2.27  & \textbf{-0.13}  & -0.15 \\ 
     \hfil{\textit{Rush hour}} &  & 59.9 & \textbf{66.4} & 2.68 & \textbf{1.98} & -0.11  & \textbf{-0.08}  \\ 
     \hfil{\textit{Touchdown pass}}  &  & 70.1  & \textbf{72.8} & 2.56  & \textbf{1.62} & -0.15 & \textbf{-0.09} \\ 
     \hfil{\textit{Snow mnt}}  &  & \textbf{65.2} & 64.6 & 2.31 & \textbf{1.04} & -0.30 & \textbf{-0.14} \\ 
     \hfil{\textit{Roller coaster}} &  & 71.1  & \textbf{73.9} & 2.21 & \textbf{1.68} & -0.15 & \textbf{-0.11} \\ 
     \hfil{\textit{Dinner Scene}} &  & 60.1 & \textbf{65.7}  & 3.91 & \textbf{1.70} & -0.09 &\textbf{-0.03}  \\ 
     \hfil{\textit{Boxing practice}} & & 65.8 & \textbf{68.6} & 2.03 & \textbf{1.50} &  -0.14 & \textbf{-0.10}  \\ 
     \hfil{\textit{Meridian}} & & 59.5 & \textbf{69.6} & 1.98 & \textbf{1.47} & -0.11 & \textbf{-0.08} \\ 
     \hfil{Average}   &   & 65.9 & \textbf{69.5} & 2.38 & \textbf{1.68} & -0.16 & \textbf{-0.12}\\  \hline
     \multicolumn{2}{|c|}{Overall average} &  65.4  & \textbf{69.7}  & 3.15 & \textbf{1.71} & -0.18 &  \textbf{-0.11} \\ \hline
  \end{tabular}
  \label{table:speed comparison}
\end{table*}

\subsection{Comparison with VP9 Speed Levels}
The VP9 encoder provides speed control settings  which allow skipping certain RDO search options and using early termination for faster encoding at the expense of RD performance. Thus, it essentially works towards the same goal as our H-FCN based partition prediction design. VP9 has 9 speed levels designated by the numerals 0-8. To use a particular speed level, the \textit{cpu-used} parameter is set to the corresponding number while encoding. The recommended speed levels for \textit{best} and \textit{good} quality settings, are 0-4, whereas levels 5-8 are reserved for use with the \textit{realtime} quality setting. Speed level 0 is the slowest and is rarely used in practice, whereas speed level 1 provides a good quality versus speed trade-off.\footnote{See \url{http://wiki.webmproject.org/ffmpeg/vp9-encoding-guide} for recommended settings.} As already pointed out earlier in this section, speed level 1 with the \textit{good} quality setting is the baseline for our method, and the fastest encoding for this configuration occurs by setting the speed level to 4.

The experimental results in Table \ref{table:performance} reveal that our method is 69.7\% faster than the baseline, with an increase of 1.71\% in BD-rate in intra-mode. However, any approach that claims to accelerate VP9 encoding is only useful in practice if it performs better than the available speed control setting. Thus, we compared the performance of our model with that of speed level 4, which is the fastest speed level of VP9 intended for this configuration (i.e. with the \textit{good} quality setting). Table \ref{table:speed comparison} summarizes the speedups, BD-rates, and BD-PSNRs of our method and those of speed level 4, where all quantities were computed with respect to the baseline.  The results in Table \ref{table:speed comparison} reveal that on the  sequences tested, our H-FCN based approach achieved a 4.3\% higher speedup on average than did the RDO at speed level 4, while incurring 1.44\% smaller BD-rate penalty. Speed level 4 suffered an average BD-PSNR decrease of -0.18 dB, as compared to -0.11 dB by our method. Our learned model delivers greater speedup, while also yielding better RD performance than the available speed control setting of the VP9 codec operating in intra-mode. The computational efficiency of the H-FCN model, combined with the effectiveness of the partitions that it predicts while maintaining good RD performance, implies that it better optimizes this important trade-off than does the VP9 speed control mechanism. 

\begin{figure}[h]
\centering
\includegraphics[width=8.5cm]{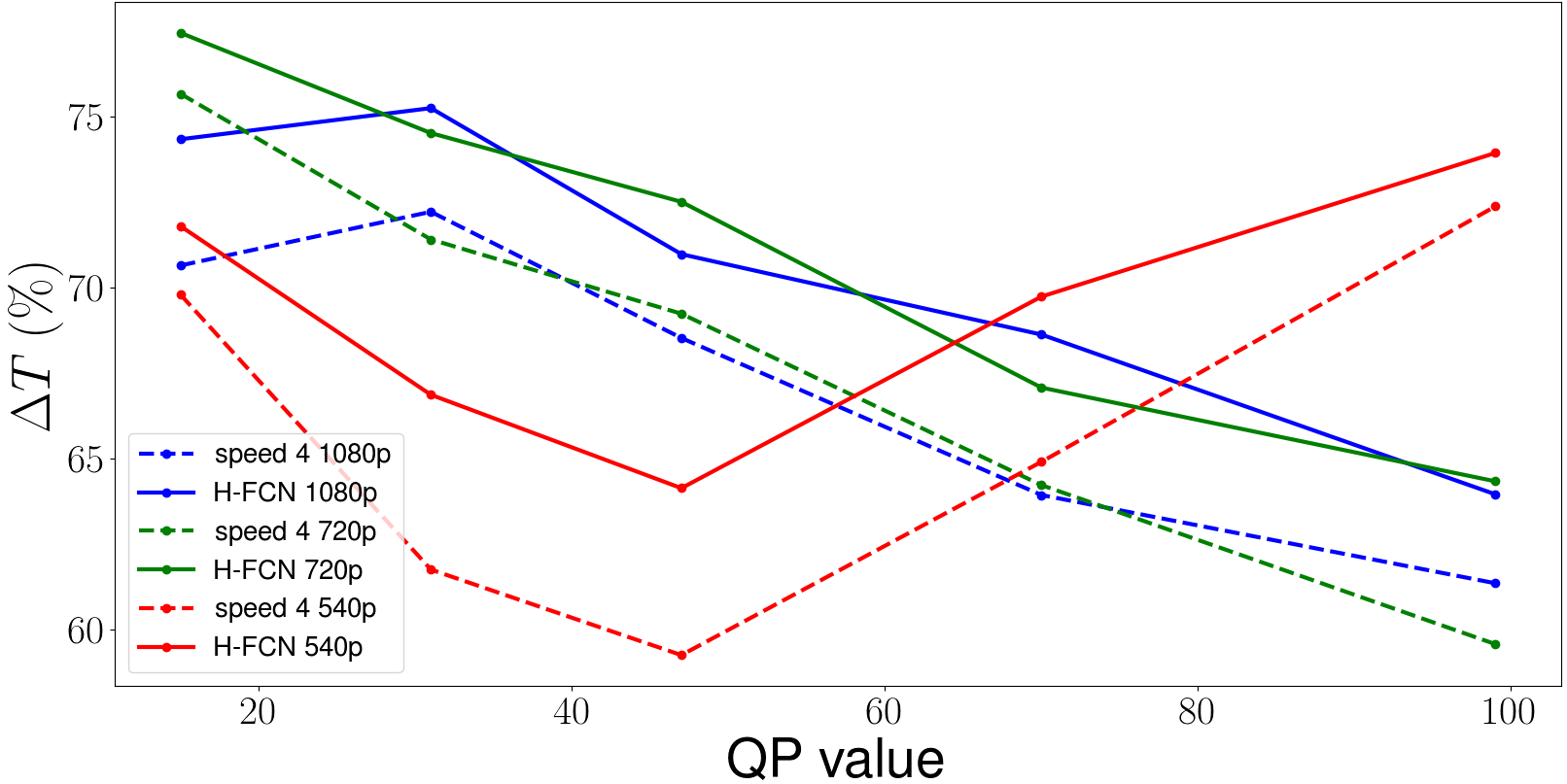}
%  \vspace{1.5cm}
\caption{\small{{Plot of speedup achieved by H-FCN and RDO at speed 4 relative to baseline against QP value, for three spatial resolutions. }}}
\label{fig:speedup}
\end{figure}

Our approach consistently  surpasses the RDO at speed level 4 in terms of speedup achieved over the common baseline, across the  practical range of QP values for adaptive streaming considered in our work. To emphasize this point further, Fig. \ref{fig:speedup} plots the net speedup achieved on the test videos against the QP values, for each of the three spatial resolutions considered. In every instance, the net speedup obtained by using the speed level 4 setting of the VP9 encoder fell below that achieved by our system, over all QP values, although at 540p resolution and QP value of 99, it was close to that of our method. This strongly supports the utility of our method for practical encoding scenarios that use a range of QP values, rather than a few specific ones.

\begin{table}[htb]
  \caption{Performance comparison of ETH-CNN \cite{inter&intra} against our approach on our test set}
  \centering
  \begin{tabular}{|p{1.9cm}|p{0.85cm}|>{\centering}m{0.85cm}>{\centering}m{0.85cm}|>{\centering}m{0.85cm}c|}
    \hline
    \multirow{2}{1.9cm}{\textbf{\hfil{Sequence}}} & \multirow{2}{0.8cm}{\textbf{\hfil{Res.}}} & \multicolumn{2}{c}{$\Delta T$ (\%)} & \multicolumn{2}{|c|}{BD-rate (\%)} \\
    % \hline
    \cline{3-6}
    &    & ETH-CNN & H-FCN & ETH-CNN & H-FCN  \\
    %\hhline{~--}
    \hline
    \hfil{\textit{Sintel trailer}}  & \hfil{\multirow{3}[20]{*}{1080p}}  & \textbf{83.1} & 74.6  & 6.40  & \textbf{5.00} \\

    \hfil{\textit{Pedestrian area}}  &   & \textbf{76.6} & 71.1 & 5.12  & \textbf{4.60} \\ 
    \hfil{\textit{Sunflower}}  &   & \textbf{79.4} & 73.9 & 2.99 & \textbf{2.65} \\ 
     \hfil{\textit{Crowd run}} &  & 54.6 & \textbf{56.1}  & 1.26 & \textbf{1.11}  \\ 
     \hfil{\textit{Ducks takeoff}} & & 58.7 & \textbf{66.7} & 2.90  & \textbf{1.49}  \\ 
     \hfil{\textit{Narrator}}  &  & \textbf{81.0} & 73.2 & 3.87 & \textbf{3.52}   \\
     \hfil{\textit{Food market}} &  & \textbf{60.0} & 59.9 & 2.02 & \textbf{1.97}   \\ 
     \hfil{\textit{Toddler fountain}}  &  & 53.4 & \textbf{57.6} & 2.15 & \textbf{1.61}  \\
     \hfil{\textit{Rainroses}}  &   & 69.8  & \textbf{70.1}  & 4.05 & \textbf{3.23} \\ 
     \hfil{\textit{Kidssoccer}}   &   & \textbf{43.7} & 41.9 & 1.14 & \textbf{1.08}  \\ 
     \hfil{Average}   &   & \textbf{66.0}  & 64.5 & 3.19 &   \textbf{2.63} \\  \hline
     \hfil{\textit{Big buck bunny}}  &  \hfil{\multirow{3}[20]{*}{720p}} & \textbf{76.8} & 71.2 & 3.21  & \textbf{2.29}  \\ 
    \hfil{\textit{Parkrun}}  &  & 47.2  & \textbf{47.5} & 1.64  & \textbf{1.43} \\ 
     \hfil{\textit{Shields}} & & \textbf{51.7} & 51.2 & 2.04 &  \textbf{1.67}  \\ 
     \hfil{\textit{Into tree}}  &  & \textbf{68.1}  & 65.9 & 1.89 & \textbf{1.35}   \\ 
     \hfil{\textit{Old town cross}}  & & \textbf{57.2} & 53.9 & 1.57 &  \textbf{1.44}  \\ 
     \hfil{\textit{Crosswalk}} &  & \textbf{75.8}  & 70.1 & 3.17 & \textbf{2.85} \\ 
     \hfil{\textit{Tango}} &  & \textbf{69.7}  & 64.8 & 3.69  & \textbf{3.52}  \\ 
     \hfil{\textit{Driving POV}} & & \textbf{53.8} & 52.8 & 1.71 &   \textbf{1.58} \\ 
     \hfil{\textit{Dancers}}  &  & \textbf{84.4}  & 74.2 & 4.96  &  \textbf{4.84}  \\ 
     \hfil{\textit{Vegicandle}} &  & \textbf{65.6} & 61.7 & 2.69  & \textbf{2.39}   \\ 
     \hfil{Average}    &  & \textbf{65.6}   &  61.7 & 2.69 &  \textbf{2.39} \\ \hline
     \multicolumn{2}{|c|}{Overall average}   & \textbf{65.8} & 63.1 & 2.94 &   \textbf{2.51} \\  \hline
  \end{tabular}
  \label{table:hevc comparison}
\end{table}

\subsection{HEVC Encoding Performance}
In this section, we assess the performance of the H-FCN model in reducing the intra-mode encoding complexity of HEVC. The H-FCN model used for this purpose was identical to the model shown in Fig. \ref{fig:HFCN} except that the first branch with 64 outputs was removed to account for the 3 level partition structure of CTUs into coding units (CUs) in the default configuration of the HM reference encoder \cite{hm} for HEVC in the intra mode. Also, size of the last convolutional layer of each of remaining three branches was changed from $1\times1\times4$ to $1\times1\times2$ since the split labels are binary for HEVC intra mode CUs. Thus, minimal changes were made to adapt the H-FCN model for HEVC partition prediction. The resulting model had  $21,418$ parameters, and was trained using the publicly available CPH-Intra database for HEVC introduced in \cite{inter&intra}. The CPH-Intra database was created using four QP values $\{22, 27, 32, 37\}$ and has $2,446,725$ and $143,925$ sample CTUs for training and validation respectively \cite{inter&intra}. The hyperparameters used for training were identical to the ones mentioned in Section \ref{subsec:hyperparams} for VP9.  Training the modified H-FCN model in the above manner enabled us to compare the performance of our model with that of the ETH-CNN model of \cite{inter&intra} trained on the same database.\footnote{the implementation of ETH-CNN as well as pretrained models are provided  at \url{https://github.com/tianyili2017/HEVC-Complexity-Reduction}} for predicting the intra mode partition of CTUs into CUs Although it might be possible to improve performance by also predicting the splits of the smallest $8\times8$ CUs into $4\times4$ prediction units (PUs), the CPH-Intra database does not include the corresponding split labels for $8\times8$ CUs, and thus our model does not make this prediction either, which allows for a fair comparison with \cite{inter&intra}.  

The accuracy values obtained by the H-FCN model on the CPH-Intra validation set were 86.90\%, 86.87\% and 91.39\% at levels 3, 2 and 1 respectively while the corresponding accuracy values for the ETH-CNN model with $1,287,189$ parameters were 90.98\%, 86.42\% and 80.42\%. The average accuracy of the H-FCN model across all the three levels was 88.38\% which was thus better than that of the much larger ETH-CNN model \cite{inter&intra}, which averaged 85.94\% across all levels. The set of test video sequences from Table \ref{table:performance} were then encoded at QP values $\{22, 27, 32, 37\}$ with the reference HEVC encoder from the HM 16.5 software  \cite{hm} using the main profile in intra mode. Each test sequence was encoded thrice, using the RDO based CTU partition decisions of the HM encoder, the partitions predicted by the ETH-CNN model of \cite{inter&intra} and our H-FCN model respectively. The 540p test sequences listed in Table \ref{table:performance} were excluded in this case, as the HM reference encoder requires the frame height to be a multiple of the minimum CU depth; since the minimum CU depth is 8 in default configuration, this condition is not satisfied for 540p sequences. 

\begin{table*}[htb]
  \caption{Performance comparison of ETH-CNN \cite{inter&intra} against our approach on the JCT-VC test set}
  \centering
  \scalebox{1}{
  \begin{tabular}{|c|c|c|cc|cc|}
    \hline
    \multirow{2}{0.85cm}{\textbf{\hfil{Class}}} & \multirow{2}{1.4cm}{\textbf{\hfil{Resolution}}} & \multirow{2}{1.4cm}{\centering \textbf{{Sequence}}} & \multicolumn{2}{c}{$\Delta T$ (\%)} & \multicolumn{2}{|c|}{BD-rate (\%)} \\
    % \hline
    \cline{4-7}
    &    &  & ETH-CNN & H-FCN & ETH-CNN & H-FCN  \\
    \hline
       \hfil{\multirow{3}[1]{*}{A}} &  \hfil{\multirow{3}[1]{*}{2560$\times$1600}} & \hfil{\textit{PeopleOnStreet}}    & 56.0 & \textbf{59.3}  & \textbf{2.37}  & 2.53 \\
         &   & \hfil{\textit{Traffic}}    & 62.8 & \textbf{64.4}  & \textbf{2.55}  & 2.63 \\
         \hline
          \hfil{\multirow{5}[1]{*}{B}} &  \hfil{\multirow{5}[1]{*}{1920$\times$1080}} & \hfil{\textit{BasketBallDrive}}    & 63.7 & \textbf{65.0}  & 4.26  & \textbf{3.60} \\
         &   & \hfil{\textit{BQTerrace}}    & \textbf{54.1} & 54.0  & 1.84  & \textbf{1.78} \\
         &   & \hfil{\textit{Cactus}}    & 59.9 & \textbf{60.9}  & \textbf{2.26}  & 2.32 \\
         &   & \hfil{\textit{Kimono}}    & \textbf{79.5} & 74.9  & 2.59  & \textbf{1.90} \\
         &   & \hfil{\textit{Parkscene}}    & 62.52 & \textbf{64.62}  & 1.96  & \textbf{1.80} \\
   \hline
    \hfil{\multirow{3}[1]{*}{E}} &  \hfil{\multirow{3}[1]{*}{1280$\times$720}} & \hfil{\textit{FourPeople}}    & 60.6 & \textbf{61.2}  & \textbf{3.11}  & 3.13 \\
    &   & \hfil{\textit{Johnny}}    & \textbf{73.2} & 67.6  & 3.82  & \textbf{3.64} \\
    &   & \hfil{\textit{KristenAndSara}}    & \textbf{70.9} & 66.1  & 3.46  & \textbf{3.26} \\
    \hline
    \multicolumn{3}{|c|}{Average}    & \textbf{64.31} & 63.80  & 2.82  & \textbf{2.66} \\
    \hline
  \end{tabular}}
  \label{table:jctvc}
\end{table*}

The comparison of the encoding performance of our approach with that of \cite{inter&intra} provided in Table \ref{table:hevc comparison} indicates that our method attains a slightly smaller speeedup than ETH-CNN on average, but is able to consistently achieve better BD-rates on all the test sequences. In this case, we performed the inference purely in Python to allow for a fair comparison with \cite{inter&intra} which does not integrate the ETH-CNN model with the HM encoder, and performs an external inference step using Python.  Integrating the trained model with the encoder using the optimized C API as we have done for VP9 should further improve the speedup achieved by our approach. The encodes used to compute the RD performance of the approach of \cite{inter&intra} were obtained using four different trained ETH-CNN models, each separately trained for one QP value, whereas our method was able to obtain better RD performance with a single H-FCN model jointly trained for all four QP values. The performance tradeoff achieved by our approach seems to be particularly promising for the 1080p sequences where 0.56\% lower BD-rate was achieved for 1.5\% increase in encoding complexity by our approach. The higher speedup attained by the approach of \cite{inter&intra} can be explained by the early termination scheme of the ETH-CNN model. On an average our approach reduced encoding time  by 63.1\% for 2.51\% increase in BD-rate, while the the method of \cite{inter&intra} resulted in 65.8\% reduction in encoding time for a 2.94\% increase in BD-rate. Using the same experimental setup that was just described, we also compared the encoding performance of our approach against that of \cite{inter&intra} on the 8 bit sequences having resolutions of at least 720p from the JCT-VC test set \cite{jct} as reported in Table \ref{table:jctvc}. Such deep learning based complexity reduction schemes are more beneficial at higher resolutions, as indicated by the lower speedup values reported on resolutions smaller than 720p in \cite{inter&intra}. Thereby, we only report the results on the highest three resolutions from the JCT-VC test set in Table  \ref{table:jctvc}. On average, our approach achieved a lower BD-rate on this set of test sequences as well, while being marginally slower than the method of \cite{inter&intra}. 

\section{Conclusion}
\label{sec:conclusion}
We have developed and explained a deep learning based partition prediction method for VP9 superblocks, that is implemented using a hierarchical fully convolutional network. We constructed a large database of VP9 superblocks and corresponding partitions on real streaming video content from the Netflix library, which we used to train the H-FCN model. The trained model was found to produce consistent partition trees yielding good prediction accuracy on VP9 superblocks. By integrating the trained H-FCN model into the VP9 encoder, we were able to show that VP9 intra-mode encoding time can be reduced by 69.7\% on average, at the cost of an increase of 1.71\% in BD-rate, by using the partitions predicted by the model, instead of invoking an RDO-based partition search during encoding. Further, by selectively employing the RDO to decide the partitions of superblocks that are predicted with inconsistent partitions, the BD-rate was reduced to 1.31\%  with a corresponding speedup of 63.4\%. The experiments we conducted comparing our model against the fastest recommended speed level of VP9 for the \textit{good} quality setting, further corroborated its effectiveness relative to the faster speed settings available in the reference encoder. This strongly suggests that the framework developed here offers an attractive alternative approach to accelerate VP9 intra-prediction partition search. We also believe that our approach is applicable to other video codecs, such as HEVC and AV1, that employ hierarchical block partitioning at multiple levels. 

Our approach can be improved to avoid the inconsistency correction step, by deploying a simple mechanism, such as early termination to enforce consistent partition predictions at all levels. Another immediate next step is to extend our approach to the prediction of  inter-mode superblock partitions, a task which encounters the additional challenge of spatiotemporal inference. Further, computing the partition tree, as well as other decisions made by the RDO, may be perceptually suboptimal, since they are generally driven by the goal of optimizing the mean squared error (MSE) with respect to the reference video. Since the MSE is generally a poor indicator of the perceptual quality  of images and videos \cite{mse}, other more perceptually relevant criteria could be considered. By instead optimizing the partition decision process in terms of a suitable perceptual quality model like SSIM \cite{ssim}, MS-SSIM \cite{ms-ssim}, VIF \cite{vif}, ST-RRED \cite{strred}, or VMAF \cite{vmaf, vmaf2}, RD performance could potentially be further improved, which is a direction that we intend to explore as part of our future work. Finally, it is also interesting to extend this approach to  optimize the AV1 codec, which, due to its even deeper partition tree, and more computationally intensive RDO search process, could benefit even more from the speedup offered by our system model. 

\section*{Acknowledgment}

We thank Anush K. Moorthy and Christos G. Bampis of Netflix for their help in creating the database. 

\begin{table*}[htb]
  \caption{Performance of the top-down and bottom-up model designs.}
  \centering
  \begin{tabular}{|p{1.2cm}|ccc|ccc|}
    \hline
     \multirow{2}{1.5cm}{\hfil{Resolution}} & \multicolumn{3}{|c|}{$\Delta T$ (\%)} & \multicolumn{3}{|c|}{BD-rate (\%)} \\
    % \hline
    \cline{2-7}
     &   Bottom-up (70,408) & Bottom-up (26,336) & Top-down (31,456) & Bottom-up (70,408) & Bottom-up (26,336) & Top-down  (31,456) \\
    \hline
   \hfil{1080p}  & 49.4 & \textbf{67.5} & 47.4 & \textbf{1.49} & 1.70  & 1.60  \\ \hline
   \hfil{720p}  & 57.8 & \textbf{72.2}& 53.3 & \textbf{1.53} & 1.75  & 1.63 \\ \hline
   \hfil{540p}  & 57.7 & \textbf{69.5} & 40.2 & \textbf{1.56} & 1.68  & 1.63  \\ \hline
   \hfil{Overall}  & 55.0 & \textbf{69.7} & 46.8 & \textbf{1.53} & 1.71  &  1.62  \\ \hline
     \end{tabular}
     \label{table:topVsbottom}
 \end{table*}
 
\appendices
\section{Comparison with Top-down Partition Prediction}

In order to substantiate the efficacy of our bottom-up hierarchical block merge prediction scheme, we designed a top-down version of the H-FCN model, to predict block splits  instead of merges, starting with the split type of the $64\times64$ superblocks. While performing top-down prediction with the hierarchical model, spatial isolation between features from the adjacent smallest blocks need to be maintained until the end of the trunk, which limits the number of max-pooling stages that can be used in the trunk. We thus modified the H-FCN architecture of Fig. \ref{fig:HFCN} such that some of the max pooling layers were removed from the trunk and instead added to the branches, so that the output of each branch is a matrix of the desired size. However, the number of convolutional filters used in a branch relative to the number of outputs predicted by that branch were the same as in Fig. \ref{fig:HFCN}. A diagram illustrating the architecture of the top-down H-FCN model is available at \cite{top_down}. The top-down model has $31,456$ parameters, which is comparable to the $26,336$ parameters in the bottom-up model from Fig. \ref{fig:HFCN} used in our work. The hyperparameters used when training the top-down model were the same as described in Section \ref{subsec:hyperparams}. 

The accuracy values of the trained top-down H-FCN model on $10^5$ samples from the validation set were  89.35\%, 82.80\%, 85.12\%, and 91.47\%, from the level 3 to level 0 of the partition tree. As compared to the corresponding values of 90.27\%,  83.47\%, 85.13\% and 91.18\% obtained using the bottom-up model with $26,336$ parameters, the accuracy improved at the lowest level, but declined at the other three levels. In fact, the reduction in accuracy was largest at the highest level of the partition tree. This is consistent with our observation that global semantics learned by the deeper convolutional layers aid the partition decisions on larger blocks. 

The encoding performance of the top-down and bottom-up partition prediction approaches are compared in Table \ref{table:topVsbottom}, where the numbers in parentheses denote the number of parameters of each model. The values in Table \ref{table:topVsbottom} were calculated by encoding the set of test video sequences listed in Table \ref{table:performance}. While the top-down model has a better BD-rate performance as compared to the bottom-up model with $26,336$ parameters, the speedup it can achieve is considerably less. However, a higher average speedup and lower BD-rate than obtained by the top-down model on average can be achieved by using a bottom-up model with $70,408$ parameters, as shown in the second and fifth columns of Table \ref{table:topVsbottom}. If the number of parameters of the top-down model is reduced, its speedup will approach the value obtained by the bottom-up model with $70,408$ parameters, but the BD-rate will increase further. This suggests that the hierarchical model is more efficient when predictions are made in a bottom-up fashion. 

The key idea behind the bottom-up partition decision is to reduce the number of computations by introducing max pooling at each stage, which progressively reduces the spatial dimensions of the feature maps that propagate along the trunk of the network. However, as the top-down approach predicts the partitions of the smallest blocks at the last stage, the number of max pooling stages that can be introduced in the trunk is limited by the need to prevent the feature maps from the adjacent smallest blocks from combining.   Hence, the prediction of the largest $8\times8$ matrix representing the split types of $8\times8$ blocks at the last branch requires a larger feature map to be carried through the trunk as compared to the bottom-up approach, which limits its speed. The results from Table \ref{table:topVsbottom} support the intuition behind adopting a bottom-up merge prediction scheme. While other frameworks can be designed that are amenable to top-down split prediction, our hierarchical model is thus found to be more suitable for bottom-up merge prediction. 

\section{Evaluation on CPH-Intra Database}

In order to evaluate the effect of the contents chosen to train the H-FCN model, we also created a VP9 partition database, from the images provided as part of the CPH-Intra database introduced in \cite{inter&intra} for HEVC intra-mode partition prediction. The CPH-Intra database is comprised of 2000 images, with 1700 images provided for training and the rest for validation and testing. As in our original database described in Section \ref{sec:database}, the contents were encoded at three different resolutions (1080p, 720p and 540p). The preprocessing stage was also the same as described in Section \ref{sec:database}, where at each encode resolution, all the images of the same or higher resolution were used as sources after Lanczos resampling (for contents that were originally at a higher resolution), and conversion to YCbCr 4:2:0 8-bit representation. However, in this case the content was a set of images, each with a distinct visual content, unlike a set of videos characterized by substantial temporal correlation between frames, as in our original database. Thus, we encoded each image individually as an intra-frame using the \textit{libvpx} reference encoder for VP9 to ensures maximal utilization of the available content. The encoding configuration, and the process of extracting superblocks from the sources and partition trees from the encodes were exactly the same as in Section \ref{subsec:database}. The final database, which we refer to as the CPH-VP9 database, consists of $841,500$ training samples. The H-FCN model trained with this CPH-VP9 database was evaluated using the set of test sequences from Table \ref{table:performance}. At each resolution, we compared the average BD-rate obtained using the CPH-VP9 database and our larger database from Section \ref{sec:database}, as reported in Table \ref{table:cph}. The average speedup is essentially the same as in Table \ref{table:performance} since the same H-FCN model was trained on both databases. 

\begin{table}[htb]
  \caption{Comparison of BD-rate on CPH-Intra database and our database}
  \centering
  \begin{tabular}{|p{1.2cm}|c|c|}
  \hline
     Resolution & CPH-VP9 (\%) & Ours (\%) \\
        \hline
   \hfil{1080p}  & 1.65 & 1.70  \\ \hline
   \hfil{720p}  & 1.58 & 1.75  \\ \hline
   \hfil{540p}  & 2.36 & 1.68   \\ \hline
   \hfil{Overall}  & 1.86  & 1.71 \\ \hline
     \end{tabular}
     \label{table:cph}
 \end{table}
As compared to the H-FCN model trained on our database, the model trained on the CPH-VP9 database achieved a lower average BD-rate on the 1080p and 720p resolutions, while its average BD-rate at 540p was higher. The overall BD-rate performance of the model trained with our database was a little better. Nevertheless, the H-FCN model was able to generalize well on the smaller CPH-VP9 database.

\balance

\end{document}